\documentclass[%
 reprint,
superscriptaddress,
nofootinbib,
 amsmath,amssymb,
 aps,
 prd,
]{revtex4-2}

\usepackage{graphicx}
\usepackage{dcolumn}
\usepackage{bm}
\usepackage{xcolor}
\usepackage{enumitem}

\usepackage[colorlinks=true,citecolor=blue,linkcolor=blue,urlcolor=blue, backref=false,pdfborder={0 0 0}]{hyperref}
\usepackage[nameinlink]{cleveref}
\usepackage{import}

\newcommand{\beq}{\begin{equation}}
\newcommand{\eeq}{\end{equation}}

\newcommand{\HEALPIX}{{\texttt{Healpix}}}
\newcommand{\metacal}{{\texttt{Metacalibration}~}}

\usepackage{xspace}

\newcommand{\LCDM}{$ \Lambda $CDM~}

\newcommand{\namaster}{\texttt{NaMaster}~}

\newcommand{\himpc}{\ensuremath{h\mathrm{Mpc}^{-1}}}

\newcommand{\photoz}{photo-$z$}
\newcommand{\bq}{\boldsymbol q}
\newcommand{\bx}{\boldsymbol x}

\newcommand{\cgg}{\ensuremath{C_{\ell}^{\gamma_{E}^{i}\gamma_{E}^{j}}}}

\begin{document}

\preprint{APS/123-QED}

\title{The Lensing Counter Narrative:\\ An Effective Description of Small-Scale Clustering in Weak Lensing Power Spectra}
\author{Joseph~DeRose}
\affiliation{Physics Department, Brookhaven National Laboratory, Upton, NY 11973, USA}
\affiliation{Lawrence Berkeley National Laboratory, 1 Cyclotron Road, Berkeley, CA 94720, USA}

\author{Shi-Fan~Chen}
\thanks{Both authors contributed equally to this work.\\
        \url{jderose@bnl.gov} \\
        \url{sc5888@columbia.edu}        
        }
\affiliation{Department of Physics, Columbia University, New York, NY 10027, USA}
\affiliation{Institute for Advanced Study, 1 Einstein Drive, Princeton, NJ 08540, USA}
\affiliation{NASA Hubble Fellowship Program, Einstein Fellow}
\date{\today}

\begin{abstract}
We present a new formalism to separate large- and small-scale contributions to cosmic shear through \textit{lensing counterterms} (LCT) inspired by effective field theory (EFT). Marginalizing over these LCTs isolates the large-scale cosmological signal in weak lensing power spectra while simultaneously constraining the impact of baryonic feedback or new physics (e.g. axion dark matter) at small scales. Our formalism removes the need for hard scale cuts in standard analyses, even when theoretical predictions are limited to below a physical cutoff $\Lambda$, resulting in significant improvements in constraining power---more than $5\times$ smaller in the case of an LSST-Y10-like analysis without marginalizing over baryons when the analysis cutoff is set to $\Lambda = 1.0h$ Mpc$^{-1}$. We conduct a proof-of-principle analysis on the publicly available DES Y3 data, finding $S_8= 0.783\pm 0.029$ and $S_8 = 0.798\pm 0.026$ for analyses with cutoffs of $\Lambda = 0.5h$ Mpc$^{-1}$ and $1.0 h$ Mpc$^{-1}$, respectively, with no detection of modifications to small-scale clustering at $k > \Lambda$ beyond the predictions of collisionless dark matter in a $\Lambda$CDM universe. We make our \texttt{JAX}-based pipeline, \texttt{gholax}, integrated with intrinsic alignment predictions from the EFT of large-scale structure at 1-loop, publicly available.
\end{abstract}

\maketitle

\section{Introduction}

Cosmic shear, the coherent distortion of galaxy shapes by weak gravitational lensing from the cosmological distribution of matter, is a powerful probe of the large-scale structure of the universe \cite{Bartelmann01,Mandelbaum18}. Already, the current generation of weak lensing surveys \cite{Amon2022a,Secco22,Dalal23,Li23,Wright25,Anbajagane25} are able to place constraints on the clustering of dark matter---frequently summarized in terms of the parameter $S_8 = \sigma_8 \sqrt{\Omega_m / 0.3}$---at the few-percent level, and imminent results from new surveys promise further, significant gains in constraining power \cite{LSST,Euclid}. Since the large-scale gravitational formation of structure can be precisely predicted from the initial conditions and particle content of the universe, these measurements enable us to put tight constraints on cosmological parameters and fundamental physics.

The precision of coming weak lensing measurements presents formidable challenges to models of the cosmic shear signal. One particular challenge is that, while structure formation can be robustly predicted on large scales where gravity dominates, the clustering of matter on small scales is significantly influenced by collisional dynamics of hot gas, often referred to as baryonic feedback, driven by galaxy formation processes such as winds from supernovae and outflows driven by active galactic nuclei. This separation of scales is critical to analyses of galaxy clustering or cross-correlations between galaxy densities and weak lensing, where well-localized samples of galaxies can be analytically modeled using perturbation theory and effective field theory (EFT) techniques on large scales without having to account for the specifics of small-scale galaxy formation \cite{Ivanov20,DAmico20,Chen22,White22}. On the other hand, weak lensing auto-correlations such as the cosmic shear signal are due to the total gravitational deflection of photons along the line-of-sight between source and observer, making the clean separation of large and small physical scales at a given angular scale difficult. Naively, the sensitivity of the cosmic shear signal to small-scale structure implies that contaminating astrophysical effects must be modeled with great care, and indeed many authors have noted the potential for their mis-modeling to bias cosmological inference \cite{Sembolini2013,Eifler2015,Chisari2019,Amon22,Arico23,GarciaGarcia2024}. Beyond purely astrophysical phenomena, many extensions to the dark sector (e.g. axion or interacting dark matter \cite{Rogers23,He23,Preston25}) also induce modifications to small-scale structure that can complicate parameter inference predicated on collisionless, cold dark matter.

While the importance of properly accounting for contributions from small-scale structure to cosmic shear is well known, its optimal solution remains a matter of significant research. The most common, and perhaps most conservative approach is to cut out the signal at affected angular scales in practical analyses; this approach is relatively intuitive but (a) requires that the amplitude and scale-dependence of small-scale astrophysics be relatively well understood despite significant unknowns remaining in the literature \cite{Schaan2020, Hadzhiyska24, McCarthy25}, and (b) requires relatively stringent cuts since the line-of-sight projection in cosmic shear makes the translation between angular and physical scales rather inexact \cite{Truttero24}. A different approach, which has seen significant progress in recent years, is to construct forward models that parameterize the impact of baryonic feedback, though this approach again relies on a reliable understanding of galaxy formation and/or sufficiently flexible parametric models \cite{Schneider15,Mead15,Mead2020,Arico20,Wayland2025}.

An alternative approach is to directly remove the contributions from small-scale structure at the level of the observable. Typically known as ``nulling'', these methods utilize linear combinations of the observed signal, canceling out contributions from small scales, thereby removing the need to model them explicitly \cite{Huterer05,Bernardeau14,Taylor2018}. Recently, ref.~\cite{Piccirilli25} proposed a nulling scheme in which small-scale modes are projected out based on their expected contamination, directly linking their nulling schemes to the expected small-scale contribution rather than relying on other metrics such as the compactness of the transformed lensing projection kernel.

In this paper, we propose yet a different track inspired by EFT techniques, constructing an effective description of small-scale matter clustering that universally describes its impact on large angular scales for all weak lensing observables via a counterterm expansion. Similar methods have recently been proposed in the context of the 1D Lyman-$\alpha$ forest power spectrum, which similarly involves an integral to small scales \cite{Garny21,Ivanov24}. For fixed assumptions about the sizes of these corrections, our method is nearly equivalent to adding a theoretical covariance to the data, i.e. constitutes a formal way to down-weight cosmic shear data more impacted by small-scale nonlinearities equivalent to an optimal nulling scheme given those assumptions. Alternatively, we show that small-scale matter clustering enters into cosmic shear observables as an extra ``noise,'' and fitting the counterterms in this effective description allows us to constrain the clumpiness of matter below the cutoff scale.

This new counterterm expansion, which introduces a compact way of marginalizing over small-scale uncertainties in weak-lensing auto-spectra, disentangles contributions to these spectra from large and small scales. This allows for a more complete extraction of information from the large scale data that can be modeled with many fewer assumptions, using theoretical tools that can utilize the separation of cosmological and astrophysical scales, while marginalizing over small scale uncertainties in a largely model agnostic manner. This is the key to understanding our counter-intuitive main result: by marginalizing over small scale physics, we are able to avoid stringent scale cuts and significantly increase the constraining power of analyses of the large scale signals of interest. 

The rest of this paper is organized as follows. In Section~\ref{sec:notation} we introduce the background notation related to galaxy lensing observables used throughout this work. Section~\ref{sec:data} describes the Dark Energy Survey Year 3 data and the measurement pipeline used to estimate angular power spectra and their covariances. Section~\ref{sec:counterterms} then presents the main theoretical advances in this work related to our lensing counterterm methodology. Section~\ref{sec:dynamical_model} describes our dynamical model, including the details of our intrinsic alignment, matter power spectrum and baryon models. Our differentiable likelihood methodology is presented in Section~\ref{sec:likelihood}, while demonstrations of our methodology on simulations and the DES-Y3 data are shown in Sections~\ref{sec:sim_results} and~\ref{sec:data_results}, concluding in Section~\ref{sec:conclusions}.

\section{Notation}
\label{sec:notation}

Throughout this paper we will be interested in the statistics of projected fields
\begin{equation}
    \delta_a(\hat{\bf n}) = \int d\chi\ W^a(\chi)\ \delta_a^{\rm 3D}(\chi \hat{\bf n}, z(\chi))
\end{equation}
where $W^a$ is the projection kernel for a given field, $a$, $\hat{\bf n}$ is the two-dimensional line-of-sight direction, $\chi$ is the comoving distance corresponding to a redshift $z$, and $\delta_a^{\rm 3D}$ is the three-dimensional field. The main observable we will be interested in in this work is the weak lensing convergence $\kappa$: the kernel for the convergence acting on a source galaxy sample $i$ is given by
\begin{equation}
    W^{\kappa^{i}}(\chi) = A (1 + z(\chi)) \chi \int_{z(\chi)}^\infty \left( \frac{ \chi(z')-\chi}{\chi(z')} \right) p^{i}(z') \ dz'
    \label{eqn:lensing_kernel}
\end{equation}
where $A = \frac32 \Omega_m H_0^2$, $p^{i}(z)$ is the galaxy redshift distribution, and the corresponding three-dimensional field is given by the nonlinear matter density field $\delta^{\rm 3D}_\kappa = \delta_m$. When computing the correlations of galaxy shapes we will also require the projected unlensed E-mode ellipticities, i.e. intrinsic alignments, of a given galaxy sample $\gamma_{E,I}^i$ with kernel $W^{\gamma_{E,I}^i} = p^i(z) E(z)$, where $E(z)$ is the normalized Hubble parameter at redshift $z$, and $\delta^{\rm 3D}_{\gamma_{E,I}^i}$ is the three-dimensional unlensed galaxy ellipticity field.

On small angular scales, or at large angular multipoles, $\ell$, the harmonic-space power spectrum of two projected fields is given by the Limber approximation \cite{Limber53,LoVerde08}
\begin{align}
\label{eq:limber}
    C^{ab}_{\ell} &= \int d\chi\ \frac{W^{a}(\chi)W^{b}(\chi)}{\chi^2} \nonumber \\
    &P_{ab}\left( k_{\perp}=\frac{\ell + \frac{1}{2}}{\chi}, k_{\parallel}=0 ; z(\chi)\right) + \mathcal{O}(\ell^{-2})\, ,
\end{align}
where $P_{ab}$ is the three-dimensional power spectrum of the unprojected fields, e.g. the nonlinear matter power spectrum $P_{\rm mm}({\bf k}; z)$ in the case of the lensing convergence. Here we have also decomposed the wavevector $\textbf{k} = (\textbf{k}_\perp, k_{\parallel})$ into its perpendicular and line-of-sight components; in the Limber approximation, projected power spectra only depend on power spectra of perpendicular modes. 

\section{Data}
\label{sec:data}
In this work, we make use of the Dark Energy Survey Year 3 \metacal shape catalog \cite{Gatti2021}, containing over 100 million galaxies with an area of 4143 square degrees and an effective number density of $\bar{n}=5.59\, \mathrm{arcmin}^{-2}$. This catalog makes use of the \metacal algorithm \cite{Huff2017,Sheldon2017}, which measures the responses, $R_{g,i}$, of the two components of galaxy ellipticities, $e_{g,i}$, to an artificial shear. These responses can then be used to weight the measured ellipticities in order to mitigate dominant shear measurement biases reducing residual biases to the $2-4\%$ level, which then must be calibrated using image simulations \cite{MacCrann2021}. We use the standard four redshift bins of this sample, which are defined and calibrated in \cite{Myles2021} using the \texttt{SOMPZ} methodology, combining wide and deep field photometric information \cite{Hartley2021} linked through the \texttt{Balrog} synthetic source injection software \cite{Everett2020,Suchyta2021} and folding in spectroscopic and high-quality multi-band photometric redshift information. These distributions are then further calibrated using clustering redshift techniques \cite{Gatti2022}. 

In this work, we are interested in modeling the angular power spectra of galaxy shapes, i.e. the harmonic space cosmic shear signal. In order to do so, we first construct maps of galaxy ellipticities using the same pipeline as in \cite{Chen24b}. We estimate the per-component shear response in each redshift bin, combining selection and measurement responses, and divide all galaxy ellipticities in the given redshift bin by these average responses. Once we have calibrated the raw ellipticities in this way, we construct ellipticity maps as
\begin{equation}
    e_{p,i} = \frac{\sum_{g\in p} v_{g} e_{g,i}}{\sum_{g\in p} v_{g}} \, 
\end{equation}
\noindent where $p$ indexes over each map pixel, $g$ indexes over galaxies in the redshift bin, and $v_g$ are the inverse variance weights provided with the \metacal catalog. Throughout this work, we use \HEALPIX\, maps with $\texttt{NSIDE}=2048$. 

We then compute the mask for each redshift bin as 
\begin{equation}
    W_{p}^{e} = \sum_{g\in p} v_{g}.
\end{equation}
Cosmic shear auto-spectra have a noise bias that must be measured and accounted for. We measure the mode-coupled noise bias as 
\begin{equation}
    N_{\ell > 2} = A_{p} \left \langle \sum_{g\in p} v_{g}^2 \sigma^2_{e,g} \right \rangle_{\rm pix}\, 
    \label{eq:nell_gammae}
\end{equation}
\noindent following \cite{Nicola2019}, where $\sigma^2_{e,g} = 0.5 (e_{g,1}^2 + e_{g,2}^2)$, $A_{p}$ is the pixel area in steradians, and the average is taken over all pixels in the map. For $\ell \leq 2$ the bias is zero as ellipticities are a spin-two field.

We then estimate harmonic space angular power spectra using the pseudo-$C_{\ell}$ algorithm as implemented in \namaster \cite{Alonso2019}. In order to invert the mode coupling matrix, we bin our $C_{\ell}$ measurements into band-powers with $\ell_{\rm min}=25$, and $\Delta \ell=3\sqrt{\ell}$. We measure all unique cross- and auto-spectra of the four DES-Y3 redshift bins, and subtract the noise bias as given in Equation~\ref{eq:nell_gammae} from the measured auto-spectra. We do not correct for the pixel window function in our measurements, as the form of the correction depends on the number of galaxies per pixel \cite{Nicola2019}, and on the azimuthal angle on the sky even in the limit of an infinite number of galaxies due to the geometry of the \HEALPIX\, pixels. At $\ell=1001$, the maximum $\ell$ fit in our analyses of DES-Y3, the fractional impact of the pixel window function is well below $1\%$, and so is negligible. Nevertheless, in future work we plan on using catalog based pseudo-$C_{\ell}$ estimators \cite{Baleato2024,Wolz2025} that avoid this complication entirely.

We use a disconnected approximation for the covariance matrix of our measurements, including mode-coupling from the survey mask through the improved narrow kernel approximation (iNKA) \cite{Nicola2019}. This has been shown to significantly improve the performance of the narrow kernel approximation (NKA) in the limit of large mode coupling from masks such as those in galaxy lensing surveys with a significant amount of small scale structure. To accomplish this, we produce an initial guess covariance matrix by evaluating our fiducial model at the Planck 2018 best fit cosmology \cite{Planck2018}, mode-coupling it using our measured mode-coupling matrices, and inserting this into the NKA as implemented in \namaster. We then find the best fit model parameters using this covariance matrix and our fiducial model and scale cuts, and regenerate a covariance matrix using the same procedure as above, now inserting the new best fit model. We neglect super-sample and connected non-Gaussian contributions to the covariance, as these have been shown to be negligible for DES-Y3 like analyses \cite{Friedrich2021}.

\section{Small-Scale Dependence of Cosmic Shear and Lensing Counterterms}
\label{sec:counterterms}

We are interested in understanding the impact of small-scale (UV) unknowns on the cosmic shear power spectrum. In particular, we have in mind a scenario where the power spectrum can be modeled arbitrarily well up to some wavenumber $\Lambda$, for example the baryonic scale $k_{\rm bar}$ in the case of dark-matter only N-body simulations or the nonlinear scale $k_{\rm NL}$ for perturbation theory, after which predictions rapidly degrade. In this case we can define the UV component of the matter power spectrum $P_{\rm mm}$ as those contributions beyond the reach of these models; as a concrete example we have
\begin{equation}
    P_{\rm mm}^{\rm UV}(k,\chi) = P_{\rm mm}(k,\chi) \Theta(k - \Lambda),
\end{equation}
where $\Theta$ is the Heaviside function, in the case of a hard time-independent UV cutoff (see Appendix~\ref{app:time_dependent_kuv} for the case of a time-dependent cutoff). For the sake of convenience, we will substitute the comoving distance $\chi$ for redshift as the time variable in the power spectrum. 

As we will now see, $P_{\rm mm}^{\rm UV}$ does not contribute arbitrarily to the cosmic shear power spectrum at large angular scales but rather does so smoothly across angular scales $\ell$ with a limited number of degrees of freedom tied to matter clustering at small scales and its evolution. This is largely due to the tapering of the lensing kernel close to the observer---nearby objects lens less efficiently than those in between the source and the observer. 

We will be interested in the behavior of the lensing kernel at low $\chi$, that is close to $z=0$.  We can construct an asymptotic expansion for this function about the origin by taking derivatives as $\chi \rightarrow 0^+$, since the lensing kernel is not defined for negative comoving distances.
Note that the integral in Equation~\ref{eqn:lensing_kernel} depends on $\chi$ in two ways: linearly in the integrand, and in the form of the lower limit at $z(\chi)$; we can re-write the latter using a Heaviside function
\begin{equation}
    W^{\kappa^{i}}(\chi) = A (1 + z(\chi)) \chi \int_{0}^\infty \left( \frac{ \chi'-\chi}{\chi'} \right) p^{i}_\chi(z') \Theta(\chi' - \chi) \ d\chi'
    \label{eqn:lensing_kernel_theta}
\end{equation}
where we have defined $p^i_\chi = H(z) p^i$ to be the source-galaxy comoving distance distribution. As $\chi \rightarrow 0^+$, derivatives of the linear part of the integrand in parentheses simply yield integrals over the whole redshift distribution, while derivatives on the Heaviside function lead to derivatives of the integrand around $\chi' = 0$.

In the limit that the source galaxy distribution is suppressed around $z=0$---specifically, when $p^{i,(n)}(\chi)$ all tend to zero at small $\chi$---we can drop the latter contributions to the low $\chi$ lensing kernel. This is formally equivalent to dropping the lower limit in Equation~\ref{eqn:lensing_kernel}, i.e. taking the limit where the lensing matter is in the foreground of all source galaxies. In this case we can write
\begin{equation}
    W^{\kappa^i}(\chi) = A\ (1 + z(\chi))\ \chi\ \left(1 - \langle \chi^{-1} \rangle_i \chi\right) \quad (\chi < \chi^i_p)
    \label{eqn:Wchi_smallp}
\end{equation}
where the expectation value is with respect to the distribution $p^{i}(z)$ and the expression is valid for comoving distances less than a characteristic LOS distance $\chi^i_p$ of the distribution $p^i(z)$. Note that while this approximation of the integral in Equation~\ref{eqn:lensing_kernel} truncates at linear order, its radius of convergence $\chi^i_p$ is finite since we're assuming that $p^{i}(z)$ drops precipitously at low $z$. It is also possible to go beyond this approximation if the low-redshift behavior of $p^i(z)$ can be well-characterized, in which case the radius of convergence gets extended  to that of the Taylor expansion of $p^i(z)$; we treat this scenario in Appendix~\ref{app:nonzero_pz}. However, it is worth noting that the tails of redshift distributions tend to be suppressed but not smooth (see e.g. Figure~\ref{fig:wmu_convergence}), making this extension potentially less applicable than it appears.

Let us now write the lensing kernel as a series
\begin{equation}
    W^{\kappa^i}(\chi) = A \sum_{n=1}^\infty w_n^{i} \chi^n
\end{equation}
where, in the approximation of Equation~\ref{eqn:Wchi_smallp}, we have
\begin{align}
    &w_1^{i} = 1, \, w_2^{i} = H_0 - \langle \chi^{-1} \rangle_i, \nonumber \\
    &w_{n>2}^i = \frac{1}{(n-1)!} \left( z^{(n-1)}_0 - (n-1) \langle \chi^{-1} \rangle_i\ z^{(n-2)}_0 \right)
\end{align}
and $z^{(n)}_0 = d^n z(\chi)/d\chi^n$ evaluated at the origin.\footnote{For convenience we list
\begin{align}
    & z^{(1)}(0) = H_0 \nonumber \\
    & z^{(2)}(0) = \frac32 \Omega_m H_0^2 \nonumber \\
    & z^{(3)}(0) = 3 \Omega_m H_0^3 \nonumber \\
    & z^{(4)}(0) = 3 \Omega_m (1 + \frac32 \Omega_m) H_0^4
\end{align}
in a flat $\Lambda$CDM universe.
} The only free parameter for each source sample in this approximation is given by the mean inverse comoving distance---indeed, it is possible to resum the derivatives of $z(\chi)$ into a redefinition of the power spectrum, as we describe in Appendix~\ref{app:alt_expansion}, in which case the series truncates past $n=2$. In general, new parameters $p^{i,(n)}(0)$ would also be required if the source distribution is not suppressed at low $z$. Importantly, the lensing kernel for any source vanishes linearly as $\chi$ nears the observer with a universal slope of $1$, tempering the effect of small-scale nonlinearities.

We can now write down the UV dependence of a lensing power spectrum between two source samples $i$ and $j$. Separating the lensing power spectrum by its contributions from $k < \Lambda$ and from UV modes $C = C_{\Lambda} + C_{\rm UV}$, and using the coordinate transformation $k = (\ell + 1/2)/\chi$ we have that
\begin{align}
    C_{\ell,\rm UV}^{ij} = \int & \frac{dk}{(\ell + 1/2)}  W^{\kappa^i}\Big(\frac{\ell + 1/2}{k}\Big) \nonumber \\
    &W^{\kappa,j}\Big(\frac{\ell + 1/2}{k}\Big) P_{\rm mm}^{\rm UV}\left(k,\chi = \frac{\ell + 1/2}{k}\right).
\end{align}
Taylor-expanding each factor in the integrand above about $z=0$ and keeping only these contributions gives
\begin{equation}
    C^{ij}_{\ell,\rm UV} = A^2 \sum_{N=2}^\infty (\ell + 1/2)^{N-1} \sum_{n+m+o=N} w_n^i w_m^j \sigma^{\rm UV}_{N,o}
    \label{eqn:lensing_ct_expansion}
\end{equation}
where we have defined the \textit{lensing counterterms} (LCTs) to be the integrated quantities
\begin{align}
    \sigma^{\rm UV}_{N,o} &= \frac{1}{o!} \int dk\ k^{-N} \left(\frac{d^o P^{\rm UV}_{mm}}{d\chi^o}\right)_{z=0} \nonumber \\
    &= \frac{1}{o!} \int_{ k > \Lambda}  dk\ k^{-N} \left(\frac{d^o P_{\rm mm}}{d\chi^o}\right)_{z=0}.
    \label{eqn:sigmas}
\end{align} 
Together with the lensing-kernel coefficients $w^i_n$, this defines a lensing-counterterm expansion for the UV dependence of the observed lensing power spectrum. Importantly, the LCTs are functionals of the matter power spectrum itself and shared between any source samples, meaning that adding more lensing observables does not result in additional free parameters in the expansion, since the lensing-kernel coefficients are functions only of source redshift distributions.

For convenience, the first few terms are
\begin{align}
    & (N=2):  \sigma_{2,0} \nonumber \\
    & (N=3):  (w_2^a + w_2^b) \sigma_{3,0} + \sigma_{3,1}  \nonumber \\
    & (N=4):  (w_2^a w_2^b + w_3^a + w_3^b) \sigma_{4,0} +  (w_2^a + w_2^b) \sigma_{4,1} + \sigma_{4,2}  \nonumber 
\end{align}
such that we gain three and six new coefficients at second order and third order, respectively. Throughout this paper, we truncate this expansion at finite $N$, in which case we will refer to the LCT expansion of $C^{ij}_{\ell,\rm UV}$ as $C^{ij}_{\ell,\rm UV,N}$.

In the LCT expansion, each subsequent term is roughly suppressed by a factor of $H_0 / \Lambda$, where the numerator comes from either the lensing-kernel coefficients or time derivative of the power spectrum, and the denominator from the integral. This suggests that the series is perturbative when the UV scale $\Lambda^{-1}$ is smaller than the Hubble radius, which gives the distance-scale over which the lensing kernel and matter power spectrum evolve. However, the radius of convergence is in fact smaller, because the lensing kernel undergoes significant evolution on scales $\chi^i_p \sim 1/\langle \chi^{-1} \rangle_i$; this implies the expansion is valid only for scales $\Lambda > \ell / \chi^i_p$, or for angular scales $\ell < \Lambda \chi^i_p$.

\begin{figure*}
    \centering
    \includegraphics[width=\linewidth]{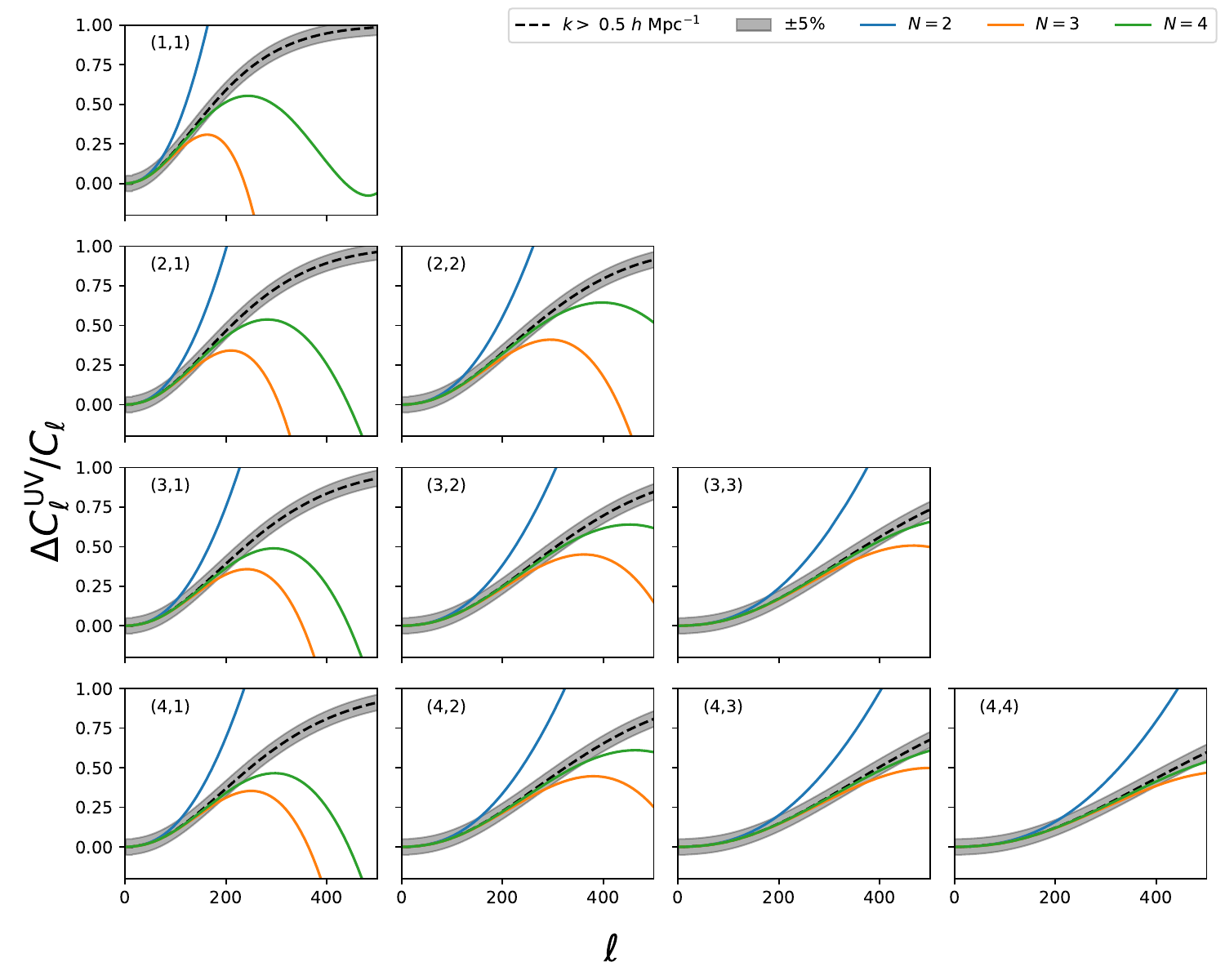}
    \caption{Lensing counterterms on the DES-Y3 cosmic shear $C_{\ell}$'s. The black dashed line shows the residual contributions due to modes shorter than $k_{\rm max} = 0.5\ h$ Mpc$^{-1}$, with $5\%$ of the total signal shown in the shaded region. These contributions parametrized via lensing counterterms (Equation~\ref{eqn:lensing_ct_expansion}), computed directly from the true nonlinear power spectrum, are shown in second, third and fourth order in blue, orange and green, showing that the expansion is perturbatively correct towards low $\ell$'s. }
    \label{fig:lensing_ct_k0p5}
\end{figure*}

\begin{figure*}
    \centering
    \includegraphics[width=\linewidth]{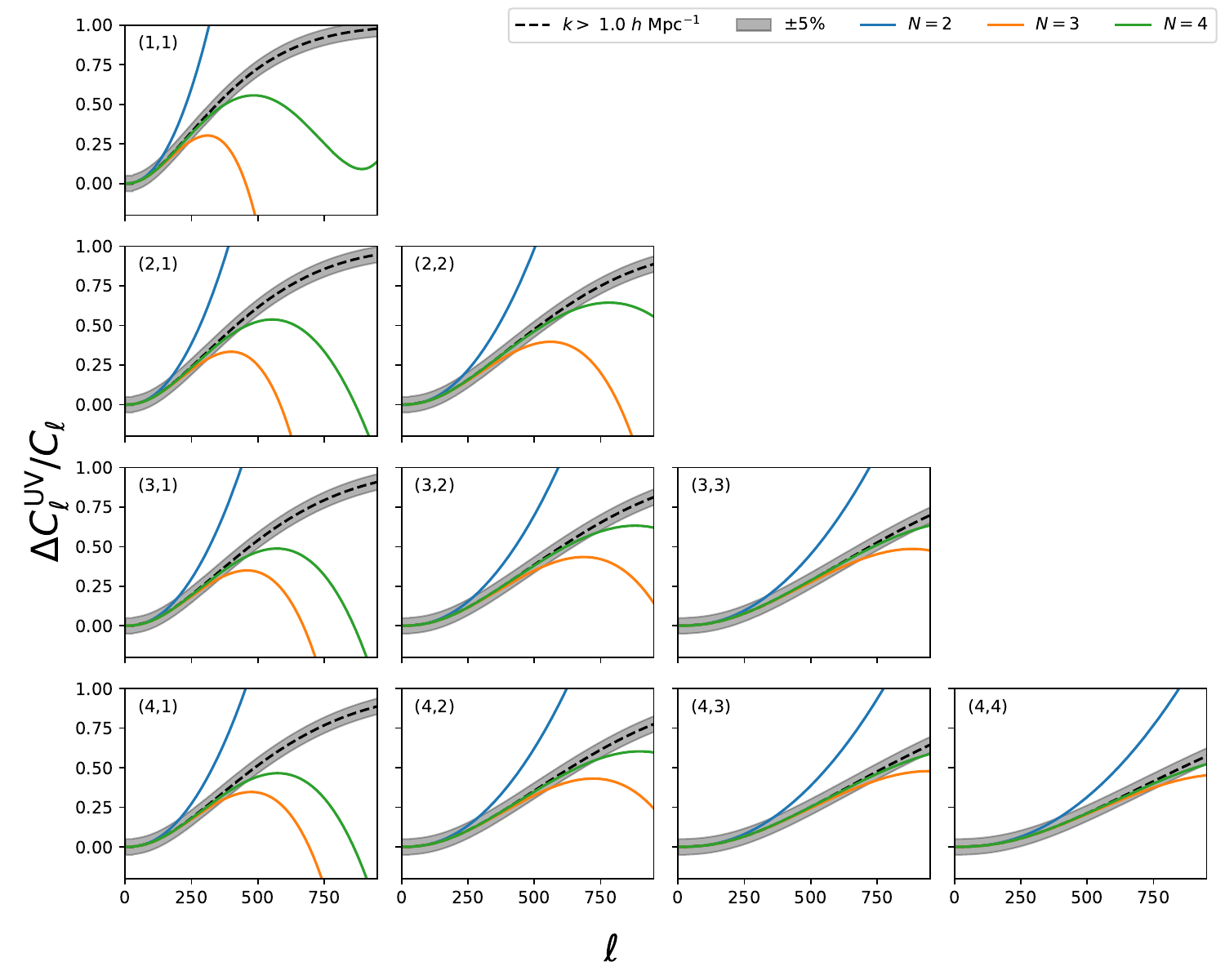}
    \caption{Like Figure~\ref{fig:lensing_ct_k0p5}, but with a cutoff of $\Lambda = 1.0\ h$ Mpc$^{-1}$.}
    \label{fig:lensing_ct_k1p0}
\end{figure*}

As an illustrative example, let us consider the case where the ``true'' nonlinear matter spectrum is given by the \texttt{HaloFit} model \cite{Smith03} as implemented in CLASS \cite{CLASS}, but where we can only robustly model it up to $\Lambda = 0.5\ h \text{Mpc}^{-1}$, for example due to baryonic effects. The black-dashed line in Figure~\ref{fig:lensing_ct_k0p5} shows the fractional contributions to the cosmic shear power spectra from scales smaller than $\Lambda$ when cross-correlating each of the four DES-Y3 source galaxy samples, with the gray band showing $5\%$ errors. These UV contributions very quickly exceed the $5\%$ error threshold when not accounted for, with the largest effects in the first bins at the lowest redshifts where the characteristic wavenumbers $\ell/\chi$ are low. The contributions from the lensing counterterms in Equation~\ref{eqn:lensing_ct_expansion}, with the counterterms computed explicitly from Equation~\ref{eqn:sigmas} rather than fit to the black-dashed curves, are shown in blue, orange and green at $N=2,3,4$. The addition of these counterterms expands the reach of the model, allowing the cosmic-shear spectra to be fit over a significantly wider range of scales with only six independent coefficients, shared across all spectra, at $N=4$. Notably, the expansion breaks down faster for the lower-redshift source bins in accordance with expectations. The case of $\Lambda = 1.0\ h \text{Mpc}^{-1}$ is shown in Figure \ref{fig:lensing_ct_k1p0}. Adopting a higher cutoff roughly doubles the reach of the LCT expansion, as expected from the scaling arguments above.

\begin{figure*}
    \centering
    \includegraphics[width=\linewidth]{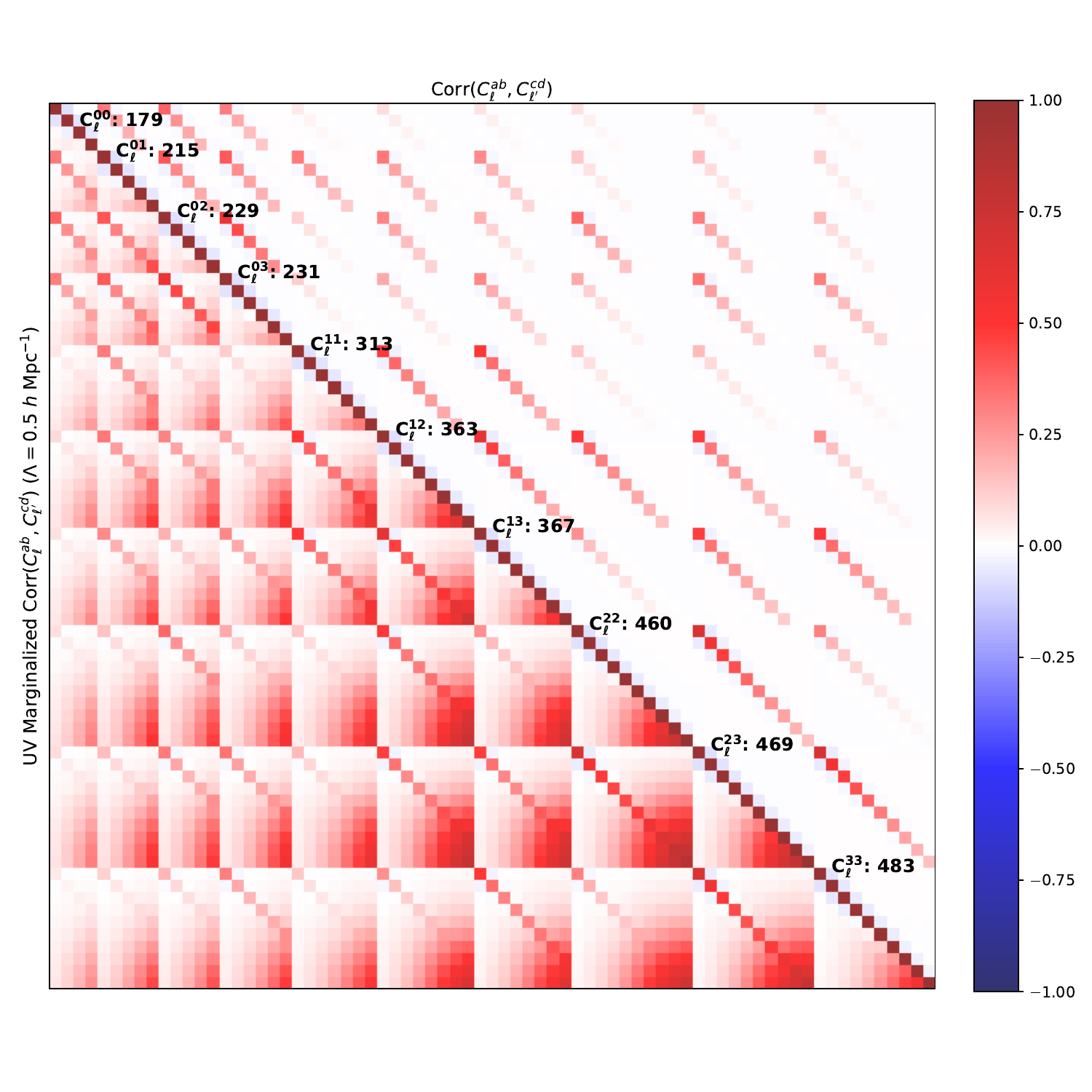}
    \caption{Correlation matrix computed for the DES-Y3 cosmic shear $C_\ell$'s with (bottom) and without (top) the implied theoretical covariance due to matter clustering below $k_{\rm max} = 0.5\ h$ Mpc$^{-1}$, assuming that matter at smaller scales deviates at the $20\%$ level relative to $N$-body simulations. Labels on the diagonal show the maximum fitted scale $\ell_{\rm max}$ in each block based on the convergence of the expansion as shown in Figure~\ref{fig:lensing_ct_k0p5}.}
    \label{fig:theory_cov}
\end{figure*}

We can alternatively interpret the lensing counterterms in Equation~\ref{eqn:lensing_ct_expansion} as a source of noise in cosmic shear spectra due to unresolved small scales. Since the lensing counterterms enter into our model predictions linearly, their contribution can be analytically marginalized to yield a data covariance matrix accounting for this small-scale noise \cite{Taylor10}. Given a Gaussian prior with the covariance $\mathcal{P}_{ab}$ over the counterterms in Equation~\ref{eqn:sigmas}, where the indices $a, b$ run over the counterterm orders $(N,o)$, the covariance matrix of the measured cosmic shear spectra is modified to
\begin{equation}
    C_{ij, \rm th} = C_{ij} + \theta^a_i \mathcal{P}_{ab} \theta^b_j
\end{equation}
where the linear templates are the derivatives of the measured cosmic shear $C_\ell$'s with respect to the lensing counterterms $\theta^a = (\partial_{\sigma_a} C^{nm}_\ell, \ldots)$ and $n, m$ are source bin pairs. This contribution adds to the effective covariance matrix along directions spanned by the polynomial LCT basis, as can be seen comparing the upper and lower parts of the implied correlation matrix in Figure~\ref{fig:theory_cov} for $\Lambda = 0.5\ h$ Mpc$^{-1}$. While the increase in the covariance seems quite significant, it is important to note that these contributions to the cosmic shear covariance are coherent, since they correspond to adding smooth polynomials $\ell^n$ to the theory prediction with fixed coefficients across bins, i.e. they do not add independently to the error bars in each cross-correlation measured. Rather, the added theoretical covariance nulls coherent changes in the measured angular correlation functions across source bins due to the same small-scale matter clustering.

\section{Dynamical Model}
\label{sec:dynamical_model}
In order to make further progress, we will need to describe the specific dynamical models we use to make predictions of weak lensing power spectra, along with the scales $k < \Lambda$ at which they are valid. In this work we are interested in the power spectrum of the projected E-mode ellipticity fields of source galaxy samples $i, j$. To leading order, the observed E-mode ellipticity field of galaxies $i$ is given by a linear combination of the  lensing convergence and the intrinsic galaxy ellipticity
\begin{equation}
\label{eqn:obs_field}
    \gamma_{E}^{i}(\hat{\bf n}) = \kappa^{i}(\hat{\bf n}) + \gamma_{E,I}^{i}(\hat{\bf n}) .
\end{equation}
While we will be primarily interested in elucidating the contributions of small-scale matter clustering to the convergence power spectra $C^{\kappa^i \kappa^j}_\ell$, we will also need to account for the auto-spectra $C^{\gamma^i_{E,I} \gamma^j_{E,I}}_\ell$ and convergence cross-spectra $C^{\kappa^i \gamma^j_{E,I}}_\ell$ of galaxy ellipticities in order to make contact with the data. In this work, our fiducial choices will be to use an $N$-body emulator and EFT-inspired parametric models of baryonic effects for nonlinear matter clustering, and the Lagrangian EFT of large-scale structure to describe galaxy shapes---we describe these in turn below.

\subsection{Nonlinear Matter Power Spectrum: $N$-body Emulator and Baryonic Feedback}
\label{sec:nonlinear_matter}

$N$-body simulations directly solve for the gravitational formation of structure down to nonlinear scales. In this work, we use the $N$-body matter power spectrum emulator\footnote{\href{https://github.com/AemulusProject/aemulus\_heft}{https://github.com/AemulusProject/aemulus\_heft}} from ref.~\cite{DeRose23} based on the \texttt{Aemulus $\nu$} suite of simulations. The \texttt{Aemulus $\nu$} emulator predicts the matter power spectrum of cold dark matter over a wide range of massive neutrino cosmologies at percent-level precision down to $\Lambda_{\rm emu} = 4 h$ Mpc$^{-1}$, using Zeldovich control variates \cite{Kokron22,DeRose23b} to reduce the effect of cosmic variance on large scales where $N$-body simulations are traditionally noisy, and transitioning to LPT at $k<0.05\himpc$. 

On top of gravitational nonlinearities, baryonic feedback plays a non-negligible role in matter clustering on the scales probed by cosmic shear. While there are a large number of hydrodynamical simulations that make predictions for this impact (e.g. \cite{LeBrun2014,Schaye2015,McCarthy2017,Pillepich2018,Springel2018,Pakmor2022,Salcido23,Schaye2023}), the precise size and scale-dependence of baryonic feedback in the matter power spectrum is still a topic of active research. In this work, we wish to remain largely agnostic to these simulations and characterize the impact of baryons in a perturbative manner, adopting the form 
\begin{equation}
    P_{\rm mm}(k) = \left( 1 - a_2 \left( \frac{k}{k_{\rm bar}} \right)^2 - a_4 \left( \frac{k}{k_{\rm bar}} \right)^4 \right) P_{\rm cdm}(k)\, .
    \label{eqn:k2n_baryons}
\end{equation}
These contributions are the leading and next-to-leading order (NLO) counterterms describing the impact of baryons on the matter power spectrum, and are expected to break down near the baryonic scale $k_{\rm bar}$ \cite{Lewandowski2014}. The baryonic scale is significantly larger than the scale of convergence for \texttt{Aemulus $\nu$}, and thus sets the UV reach of our matter power spectrum prediction $\Lambda \lesssim k_{\rm bar}$. We note that while we adopt the \texttt{Aemulus $\nu$} model for $P_{\rm cdm}(k)$, in principle one could replace this with a perturbative prediction for $P_{\rm cdm}(k)$, in which case the nonlinear scale would instead set $\Lambda \lesssim k_{\rm nl}$.
 
In addition to this form, we also make use of an alternative parameterization of baryonic physics, opting to include a Pad\'e approximant to the leading order counterterm rather than include the next-to-leading order $k^4$ term \cite{Sullivan21}
\begin{equation}
    P_{\rm mm}(k) = \left( 1 - a_2  \frac{(\tilde{R}\, k/k_{\rm bar})^2}{1+(\tilde{R}\, k/k_{\rm bar})^2} \right) P_{\rm cdm}(k)\, .
    \label{eqn:pade_baryons}
\end{equation}
Taylor expanding this expression recovers the $k^4$ contribution above, but this form also re-sums higher order terms. This alternative form extends the range of validity of our model for baryonic physics when compared to hydrodynamical simulations, as demonstrated below, at the expense of being more phenomenological.

In order to determine the range of validity of these perturbative descriptions of baryonic effects we have compared them to the \texttt{SP(k)} model \cite{Salcido23}, which fits a wide range of hydrodynamical simulations. We make use of this model as a point of comparison as it provides a simple prescription for predicting a broad range of baryonic scenarios, from the low to high feedback regime. That being said, it is tuned to simulations run with one simulation code and so we perform further spot checks of our models against widely known hydrodynamical simulations in Section~\ref{subsec:noiseless_tests}.

\begin{figure*}
    \centering
    \includegraphics[width=\linewidth]{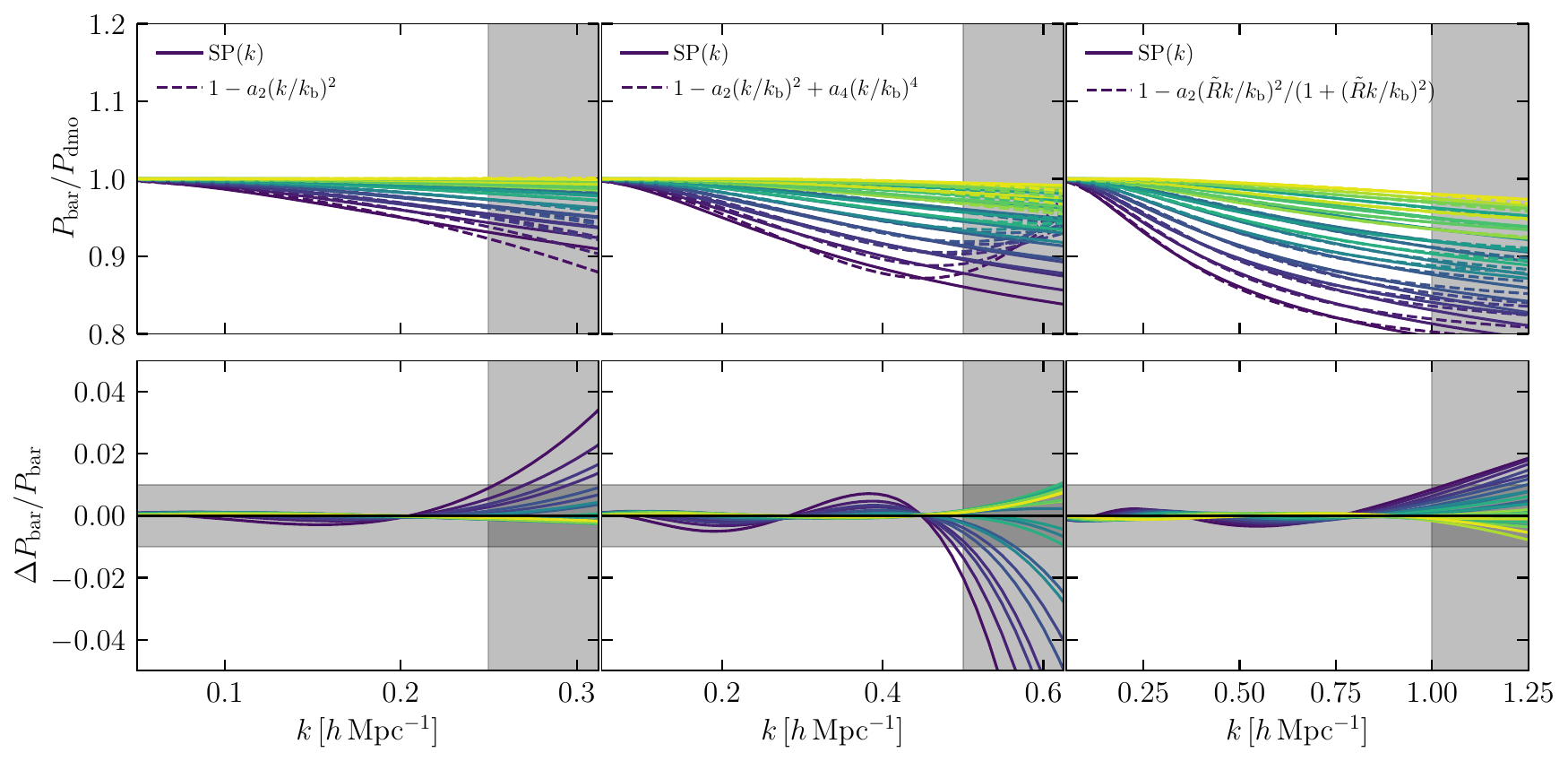}
    \caption{({\it Top}) Comparison of different parameterizations of baryonic feedback on the matter power spectrum (dashed) to a range of hydrodynamical simulation predictions as computed by the $\textrm{SP}(k)$ model \cite{Salcido23} (solid). ({\it Bottom}) Fractional residual of the baryonic feedback parameterizations used in this work from the hydrodynamical predictions. The first two columns show fits of counterterm expansions truncated at second and fourth order, which work at the percent level to $k_{\rm max}= 0.25\, \himpc$ and $k_{\rm max}= 0.5\, \himpc$ respectively. The rightmost panel adds a Pad\'e approximant to the second order expansion to control high-$k$ behavior, extending the validity of this model to $k_{\rm max}= 1\, \himpc$ at the cost of enforcing that the leading order and next-to-leading order contributions have the opposite sign.}
    \label{fig:baryon_models}
\end{figure*}

Figure~\ref{fig:baryon_models} shows the results of these comparisons at $z=0.125$, the lowest redshift modeled by \texttt{SP(k)}, where the three columns show models with progressively increasing levels of complexity starting with the standard $k^2$ counterterm on the left, adding an additional $k^4$ term in the middle column, and finally showing the $k^2$ term with a Pad\'e approximant on the right hand side. The top row shows a comparison of the perturbative models as dashed lines to \texttt{SP(k)} models with varying baryonic feedback strengths as solid lines where the various \texttt{SP(k)} models are the same across all panels, and darker colors represent stronger baryonic feedback. For these fits, and throughout this work, we take $k_{\rm bar} = 1.25\, \himpc$ which implies a $10\%$ ($1\%$) contribution to $P_{\rm mm}(k)$ at $k=0.4\himpc$ for $a_{2}=1$ ($a_{4}=1$). We find that the three different perturbative models shown here are able to fit \texttt{SP(k)} predictions to $k=0.25\, , 0.5\, , \textrm{and }1.0\, \himpc$ at the $\sim1\%$ level respectively, as represented by the gray regions in the top row of panels. Given the empirically-motivated functional form of the \texttt{SP(k)} model, we don't expect our models to fit it better than this level even though they are by construction asymptotically correct at low $k$. Recently, \cite{Bartlett2025} also investigated the ability of similar models to fit a range of hydrodynamical prescriptions, finding comparable levels of accuracy.

Furthermore, the values of $a_2$ and $a_4$ preferred by these fits are roughly covered by unit-variance Gaussian distributions, motivating our use of this distribution as a prior for these terms. The distribution of $\tilde{R}$ preferred by these simulation fits is broader and peaked around $\tilde{R}\sim1$, but is conservatively covered by a Gaussian distribution with standard deviation of 2, motivating our prior on this parameter. 

In order to model the redshift dependence of baryonic effects, we allow these parameters to evolve as linear splines in redshift
\begin{align}
    X(z)= \sum_{m=0}^{N} X_{m} W\left( \frac{z - z_{\rm min}}{\Delta} - m \right)
    \label{eqn:spline}
\end{align}
where $X\in \{a_2, a_4, R\}$ and $\Delta$ is a pre-set redshift spacing defining the smoothness of the redshift dependence and the spline covers points between $z_{\rm min}$ and $z_{\rm max} = z_{\rm min} + N \Delta$. As fiducial choices we take $N=3$, and $\Delta=0.6$, i.e. the baryonic effects evolve piecewise-linearly with one node between $z=0$ and $z=1.2$.

\subsection{Intrinsic Alignments}

We make use of the Lagrangian EFT of galaxy shapes developed in ref.~\cite{Chen24} to model intrinsic alignment cross- and auto-power spectra. Following ref.~\cite{Vlah20}, the three-dimensional shapes of galaxies can be decomposed into a spherical harmonic basis with angular momentum $\ell = 2$ and helicities $|m| \leq 2$, with nonzero cross correlation only at equal helicities (noting that galaxy densities are $\ell, m = 0$ scalar fields). The helicity spectra are related to the observed, two-dimensional projected fields in the Limber approximation via \cite{Vlah21}
\begin{align}
    &P_{\gamma^i_{E,I} \kappa^j}(k,z) = \sqrt{\frac{3}{8}} P_{02, ij}^{(0)}(k,z), \nonumber \\
    &P_{\gamma^i_{E,I} \gamma^j_{E,I}}(k,z) = \frac38 P_{22,ij}^{(0)}(k,z) + \frac{1}{8} P_{22,ij}^{(2)}(k,z)
\end{align}
where $P_{\ell \ell',ab}^{(m)}$ are the cross power spectra between the $\ell$ and $\ell'$ components of the galaxy shape ($i,j$) and matter ($m$) field with helicity $m$. 

Analytic predictions for these helicity spectra are implemented in the publicly available \texttt{Python} code \texttt{spinosaurus}\footnote{\url{https://github.com/sfschen/spinosaurus}}. Briefly, in the Lagrangian EFT the (number-density weighted) galaxy shape field is described by a bias functional $g_{ij}$ at their initial, Lagrangian positions $\bq$ and advected to their observed positions via displacement $\Psi$, i.e.
\begin{equation}
    g_{ij}(\bx) = \int d^3\bq \ \delta_D(\bx - \bq - \Psi(\bq))\ g_{ij}(\bq).
\end{equation}
The bias functional $g_{ij}$ can be constructed order-by-order out of operators constructed from the local initial conditions of galaxies, and is given up to 1-loop order by
\begin{align}
    g_{ij}(\bq) &=  \, A_{1} s_{ij} + A_{\delta1}\delta s_{ij}  + A_{2} \mathrm{TF}\{s^2\}_{ij} + A_{t} t_{ij}  \nonumber \\
 & + A_{\delta t} \delta t_{ij} + A_{3} \mathrm{TF}\{L^{(3)}\}_{ij} + \alpha_{s}\nabla^2 s_{ij} + \epsilon_{ij}\, 
  \label{eqn:shape_exp}
\end{align}
where $s_{ij}$ and $\delta$ are the initial tidal shear and matter overdensity and $\epsilon_{ij}$ corresponds to the stochastic shape noise. The commonly used nonlinear alignment (NLA) and tidal-alignment and tidal-torquing (TATT) models of intrinsic alignments \cite{Blazek19} roughly correspond to allowing $A_1$ and $A_1, A_{\delta 1}, A_2$ to be nonzero, respectively, though we note that the latter does not include counterterms like $\alpha_s$ and $\epsilon_{ij}$  required to regularize the unknown small-scale dependence of the theory; we refer the interested reader to ref.~\cite{Chen24} for the definitions of the other operators. In this work we keep the dimensionless bias coefficients in Eq.~\ref{eqn:shape_exp} up to second order, consistent with what was done in \cite{Chen24b}, since the two independent third-order contributions are rather degenerate with the counterterm. In addition, we have subtracted a fixed stochastic contribution from our measurements corresponding to the expectation given in Section~\ref{sec:data}, noting that contributions to the stochasticity from perturbative modes are very small in comparison. We leave modeling the IA contribution to cosmic shear with sample-dependent EFT parameters, including the stochasticity, to future work.
 
Finally, we parametrize the sample and time dependence of the bias operators in equation~\ref{eqn:shape_exp} as a two-parameter spline as in Equation~\ref{eqn:spline}. We have adopted this form to have degrees of freedom for IA evolution consistent with other state-of-the-art cosmic shear analyses in order to focus specifically on the impact of small-scale matter clustering in this work, though we caution that more general parameterizations may be required for future surveys \cite{Chen24b}. We intend to return to this topic in a future work.

\section{Likelihood and Sampling}
\label{sec:likelihood}

\subsection{Parameters and Priors}
\label{sec:params}
\begin{table*}
\caption{Parameters and priors}
\begin{center}
\begin{tabular}{| c  c  c  |}
\hline
\hline
Parameter & Prior & Reference\\  
\hline 
\multicolumn{3}{|c|}{{\bf Cosmology}} \\
$\omega_c$  &  $\mathcal{U}$($0.08, 0.16$) &  \\ 
$A_s$ &  $\mathcal{U}$($1.1\times 10^{-9},3.1\times 10^{-9}$) & \\
$n_s$ & $\mathcal{U}$($0.93, 1.01$) & Sec. \ref{sec:likelihood} \\
$\omega_b$ & $\mathcal{U}$($0.0173, 0.0272$) &  \\
$h$  & $\mathcal{U}$($0.52, 0.82$) & \\
$\sum \, m_{\nu}$  & 0.06 eV & \\
\hline
\multicolumn{3}{|c|}{{\bf Lensing counterterms}} \\
$f^{UV}_{N,o}$   & $\mathcal{N}$ ($0,0.4^2$) & Eq. \ref{eqn:sigmas}\, , \ref{eqn:lct_param}\\
\hline
\multicolumn{3}{|c|}{{\bf Baryonic Effects}} \\
$a_{2,i}$   & $\mathcal{N}$ ($0,1^2$) & Eq. \ref{eqn:k2n_baryons}\, , \ref{eqn:spline}\\
$a_{4,i}$   & $\mathcal{N}$ ($0,1^2$) & Eq. \ref{eqn:k2n_baryons}\, , \ref{eqn:spline}\\
$\tilde{R}_{i}$   & $\mathcal{N}$ ($0,2^2$) & Eq. \ref{eqn:pade_baryons}\, , \ref{eqn:spline}\\
\hline
\multicolumn{3}{|c|}{{\bf Intrinsic Alignment}} \\
$(\frac{\sigma_8(z)}{\sigma_{8,\rm fid}})c_{1}^{i}$   & $\mathcal{N}$ ($0,5^2$) & Eq. \ref{eqn:spline}\, , \ref{eqn:shape_exp}\\
$(\frac{\sigma_8(z)}{\sigma_{8,\rm fid}})^2c_{2}^{i}$   & $\mathcal{N}$ ($0,5^2$) & Eq. \ref{eqn:spline}\, , \ref{eqn:shape_exp}\\
$(\frac{\sigma_8(z)}{\sigma_{8,\rm fid}})^2c_{\delta 1}^{i}$   & $\mathcal{N}$ ($0,5^2$) & Eq. \ref{eqn:spline}\, , \ref{eqn:shape_exp}\\
$(\frac{\sigma_8(z)}{\sigma_{8,\rm fid}})^2c_{t}^{i}$   & $\mathcal{N}$ ($0,5^2$) & Eq. \ref{eqn:spline}\, , \ref{eqn:shape_exp}\\
$(\frac{\sigma_8(z)}{\sigma_{8,\rm fid}})^2\alpha_{s}^{i}$   & $\mathcal{N}$ ($0,45^2$) & Eq.  \ref{eqn:spline}\, , \ref{eqn:shape_exp}\\
\hline
\multicolumn{3}{|c|}{{\bf Shear calibration}} \\
$m^{i}$ & $\mathcal{N} \text{(Table \ref{tab:source_info})}$ & Eq. \ref{eqn:shear_calib} \\
\hline 
\multicolumn{3}{|c|}{{\bf Source \photoz\ }} \\
$\Delta z^{i}_{\rm s}$  & $\mathcal{N}$ ($0.000$, \text{Table \ref{tab:source_info}}) & Eq. \ref{eqn:delta_z_source}  \\
\hline
\end{tabular}
\end{center}
\label{tab:params}
\end{table*}

We make use of a Gaussian likelihood approximation for all of our analyses, with priors detailed in Table~\ref{tab:params}. We assume a \LCDM model with flat priors on the cosmological parameters, with bounds set by the extent of the parameter space used to generate the simulations that were used to train the Aemulus $\nu$ matter power spectrum emulator \cite{DeRose23}. We have fixed $\sum\, m_{\nu}$ to the minimal mass allowed by the normal hierarchy of neutrino masses in order to reduce parameter projection effects. We set priors on our baryonic feedback parameterization as described in Section~\ref{sec:nonlinear_matter}, based on the range of these parameters found in a broad range of hydrodynamical simulations. 

We also assume wide priors on our intrinsic alignment parameters, where the standard deviations of these parameters are set to be close to what has been found for halos with mean masses of approximately $M=10^{12}\, h^{-1}M_{\odot}$ \cite{Akitsu2023}, a conservative choice given that in all simulations to date the alignments of galaxies have been found to be many times smaller than that of their host halos. We note that these priors are narrower than those in \cite{Chen24b} by a factor of two, which we had previously inserted into our normalization in order to make contact with previous literature which was normalized with respect to the projected rather than three-dimensional alignment amplitude. 

In addition to the free dynamical parameters above we will marginalize over the lensing counterterms in \S\ref{sec:counterterms} directly. We take advantage of the fact that our model for the matter power spectrum does not break down instantaneously at $\Lambda < \Lambda_{\rm emu}$ in order to parametrize
\begin{equation}
    \sigma^{\rm UV}_{N,o} = \sigma^{\rm UV, \rm fid}_{N,o}f^{\rm UV}_{N,o}
    \label{eqn:lct_param}
\end{equation}
\noindent 
where we compute the cosmology-dependent $\sigma^{\rm UV, \rm fid}_{N,o}$ using Eq.~\ref{eqn:sigmas}, plugging in our fiducial dark--matter--only matter power spectrum, $P_{\rm cdm}(k)$. In this way the $f^{\rm UV}_{N,o}$ are left as free parameters characterizing the fractional difference from the fiducial UV contributions. Note that although we have chosen to keep contributions between $\Lambda$ and $\Lambda_{\rm emu}=4\, \himpc$ in our fiducial model, the lensing counterterms marginalize over uncertainty in this range as well as in the range $k>\Lambda_{\rm emu}$. 

In order to compute $\sigma^{\rm UV, \rm fid}_{N,o}$, we use first-order one-sided finite differences with $\Delta z = 0.01$ to compute the derivatives in Equation~\ref{eqn:sigmas}, in order to avoid potential inaccuracies in our emulator at the edge of our training set, i.e. $z=0$. Varying the order of the finite difference calculation leads to $\sim2-3\%$ variations in $\sigma^{\rm UV, \rm fid}_{N,o}$, but we are not particularly sensitive to these given the wide priors on $f^{\rm UV}_{N,o}$ that we assume. When computing scale cuts as described in the next section, we have verified that fits of $f^{\rm UV}_{N,o}$ to $C^{ij}_{\ell, \rm UV}$ produce $\ell_{\rm max}$ values for all bin combinations that are within $5\%$ of our fiducial implementation.

\subsection{Scale cuts}
\label{sec:scale_cuts}
One of the main advantages of our LCT methodology is the ability to extract $k<\Lambda$ information from angular scales that are significantly contaminated by $k>\Lambda$. Our ability to do this while delivering unbiased cosmological constraints is a strong function of the order of the LCT expansion. As such, we must set different scale cuts depending on the LCT order that we expand to. In order to do this, we require that the LCT expansion be converged to some fractional accuracy, $f$, i.e.
\begin{equation}
    \frac{C^{ij}_{\ell, \rm UV}-C^{ij}_{\ell, {\rm UV}, N}}{C^{ij}_{\ell}} < f\, ,
\end{equation}
\noindent where $C^{ij}_{\ell, {\rm UV}, N}$ is given by Equation~\ref{eqn:lensing_ct_expansion}, truncated at order $N$ and is zero for $N=1$. For the DES-Y3 analyses below (including simulated analyses), we take $f=0.05$, i.e. we make scale cuts where the LCT expansion breaks down at the $5\%$ level compared to the total signal. This is analogous to how \cite{Doux2021} set scale cuts for a given $k_{\rm max}$. For the LSST-like simulated analyses below we apply a more stringent $1\%$ requirement given the significantly smaller statistical errors. For DES-Y3 we make low-$\ell$ scale cuts at $\ell=50$, as below this there is evidence for B-modes in the data \cite{Chen24b}, and we do not expect significant gains from low $\ell$ data with DES. For LSST we remove this low-$\ell$ cut. We use these scale cut criteria throughout except where explicitly stated otherwise.

\subsection{Shear Calibration and Redshift Distribution Uncertainties}
We implement standard observational systematic marginalization models in order to account for imperfect calibration of the shapes and redshift distributions provided with observed shear catalogs. In order to account for uncertainties in the calibration of shapes, largely due to blending of galaxies \cite{Gatti2021,MacCrann2021}, we include a multiplicative bias term for each redshift bin 

\begin{equation}
\label{eqn:shear_calib}
    \cgg \to (1 + m_{i})(1 + m_j) \cgg\, .
\end{equation} 

To model the uncertainties on the redshift distributions of the source samples, we make the common assumption that these uncertainties can be parameterized as shifts in the mean redshift of each bin, $\Delta z_i$. We then use the appropriately shifted redshift distributions,
\begin{equation}
    p^{i\prime}(z) = p^{i}(z + \Delta z_i)\, ,
\label{eqn:delta_z_source}    
\end{equation}
\noindent to calculate angular power spectra. The multiplicative biases and redshift distribution uncertainties that we use for the DES-Y3 analyses performed in this work are listed in Table~\ref{tab:source_info}. For the LSST-Y10-like analyses done below, we assume mean multiplicative biases of zero, with uncertainties of $\sigma(m)=0.003$, and uncertainties on the mean redshifts of $\sigma(\Delta z)=0.001$, following the LSST Dark Energy Science Collaboration's Science Requirements Document \cite{Mandelbaum2021}. 

\begin{table}[htb!]
    \centering
    \begin{tabular}{|c|c|c|c|c|}
         \hline
         \hline
         Source bin & $\sigma \left(\Delta z_{s}\right)$ & $\langle m \rangle$ & $\sigma (m)$ \\
         \hline
         0 & 0.018& -0.006& 0.009\\
         1 & 0.015& -0.020& 0.008\\
         2 & 0.011& -0.024& 0.008\\
         3 & 0.017& -0.037& 0.008\\
         \hline 
    \end{tabular}
    \caption{Mean redshift uncertainties, $\sigma \left(\Delta z_{s}\right)$, multiplicative biases, $\langle m \rangle$, and their associated uncertainties $\sigma (m)$ for each DES-Y3 redshift bin.}
    \label{tab:source_info}
\end{table}

\subsection{Differentiable Emulators and Sampling}
In order to accelerate our analyses, we have trained multi-layer perceptron (MLP) emulators for the theory calculations that are the bottlenecks of our likelihood calculations, i.e. Boltzmann code and perturbation theory calculations. The architecture of these MLPs is similar to those we have used in previous works \cite{DeRose2021,Chen24b}, except that here we train a single emulator to predict all basis spectra of our intrinsic alignment model at once, and that we predict spectra that are normalized by $D(z)^2$ to remove the dominant redshift evolution of the spectra similar to \cite{Bonici25}. See Appendix~\ref{app:emu_construction} for more details. 

Another major advantage of using MLPs as emulators for relatively slow theory codes is that calculations implemented using MLPs, being simple matrix multiplications followed by the application of non-linear activation functions, are trivially differentiable. Thus, when embedded in a likelihood code that is also differentiable as we have done in this work, these emulators enable the calculation of the derivatives of our model predictions and likelihood with respect to the model parameters. This is useful for a number of reasons, but in this work the main application we make of this differentiability is the use of Hamiltonian Monte Carlo (HMC) sampling. The HMC algorithm is a variant of the Metropolis-Hastings algorithm where proposals are generated by running Hamiltonian simulations with the potential energy of the Hamiltonian system given by the negative log-likelihood \cite{Simon1987}. In this way, proposals can be generated with high acceptance fractions, and low correlations with previous samples such that significantly fewer samples are required to achieve converged estimates of the posterior distribution of parameters. 

In order to take advantage of these improvements, one must take gradients of the Hamiltonian, i.e. the log-likelihood, with respect to the model parameters. Estimating these gradients with finite difference methods is noisy and slow, so instead we write our likelihood code in the automatically differentiable \texttt{JAX} code. In addition to making our likelihood automatically differentiable, this also allows us to just-in-time compile our calculations and seamlessly run them on GPUs, leading to order of magnitude speed-ups over non-JIT compiled code run on CPUs. We release our \texttt{JAX}-powered likelihood code, called \texttt{gholax}\footnote{\url{https://github.com/j-dr/gholax}}, with this paper.

In this work, we use the No U-Turn Sampling (NUTS) variant of HMC \cite{Hoffman2014} as implemented in the \texttt{Blackjax} library \cite{cabezas2024}. In order to facilitate numerically stable gradients of our model with respect to our parameters, we normalize all parameters by the standard deviation of their priors in our code. Before sampling, we run an L-BFGS minimization \cite{Liu89} on the negative log-posterior in order to start our sampling from the best fit value, which significantly speeds up the warm-up phase of our sampling by enabling an accurate starting point for the HMC mass matrix, which we assume to be diagonal. In order to estimate the mass matrix and step size used in the Hamiltonian integration, we run four independent warm-up phases in parallel for 500 steps and use the median mass matrix and step size in further sampling. We then run four NUTS chains with the mass matrix and step size estimated from this warm-up phase, continuing to sample until we measure Gelman-Rubin statistics \cite{Gelman1992} on all two-dimensional parameter subspaces to be $R-1< 0.1$.

\section{Results on Simulations}
\label{sec:sim_results}
\subsection{Fitting Noiseless Mock Data}
In order to demonstrate the utility of the lensing counterterms introduced in this work, we first fit noiseless mock data generated with our fiducial model without counterterms, including contributions to $\Lambda_{\rm emu} = 4\, \himpc$. In particular, we generate DES-Y3-like and LSST-Y10-like noiseless simulations. The DES-Y3-like data is generated using the DES-Y3 fiducial source galaxy redshift distributions \cite{Myles2021}, assuming our fiducial model evaluated at the Planck 2018 cosmology \cite{Planck2018} with all nuisance parameters set to zero. The LSST-Y10-like data assumes five source redshift bins, following the set up described in \cite{Fang2023}, evaluated at the same cosmology with nuisance parameters set to zero. For the DES-like analyses, we use the Gaussian covariance described in Sec.~\ref{sec:data}, while for the LSST-like analyses we produce Gaussian covariance matrices, assuming $f_{\rm sky}=0.4$, $\sigma_e=0.26$, and $\bar{n}_{\rm eff}=27.5\, \textrm{arcsec}^{-2}$, similar to what is assumed in \cite{Mandelbaum2021}. 

First, we explore how the constraining power of our mock DES-Y3 and LSST-Y10 analyses varies as we increase the order of the lensing counterterm expansion. For these analyses, we take $\Lambda=1 \himpc$ and fix baryon and intrinsic alignment contributions to zero. As we increase the order of the LCT expansion, we expand the range of scales that we include in our analyses following the criteria described in Section~\ref{sec:scale_cuts}.

\begin{figure*}
    \centering
    \includegraphics[width=0.45\linewidth]{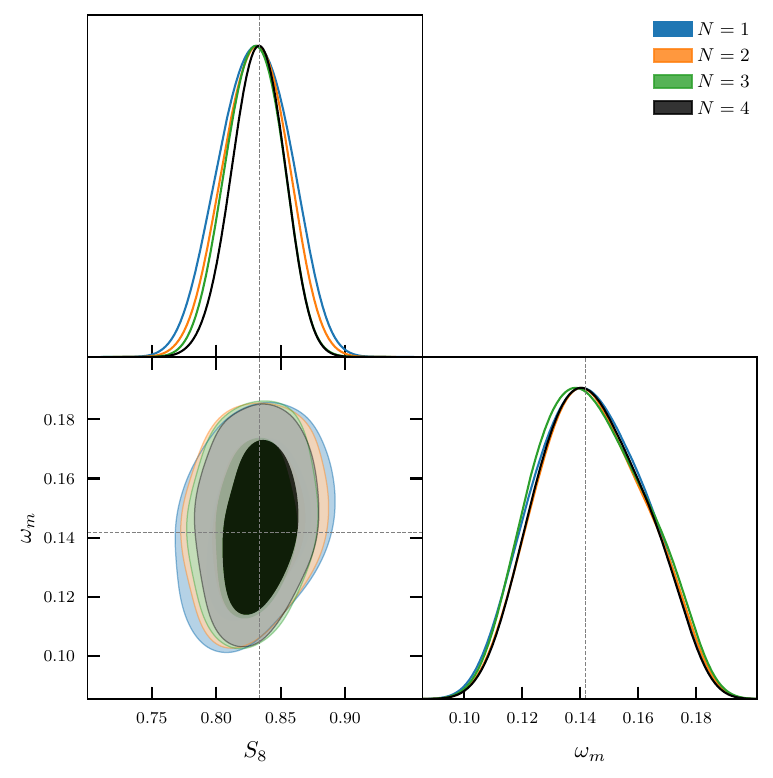}
    \includegraphics[width=0.45\linewidth]{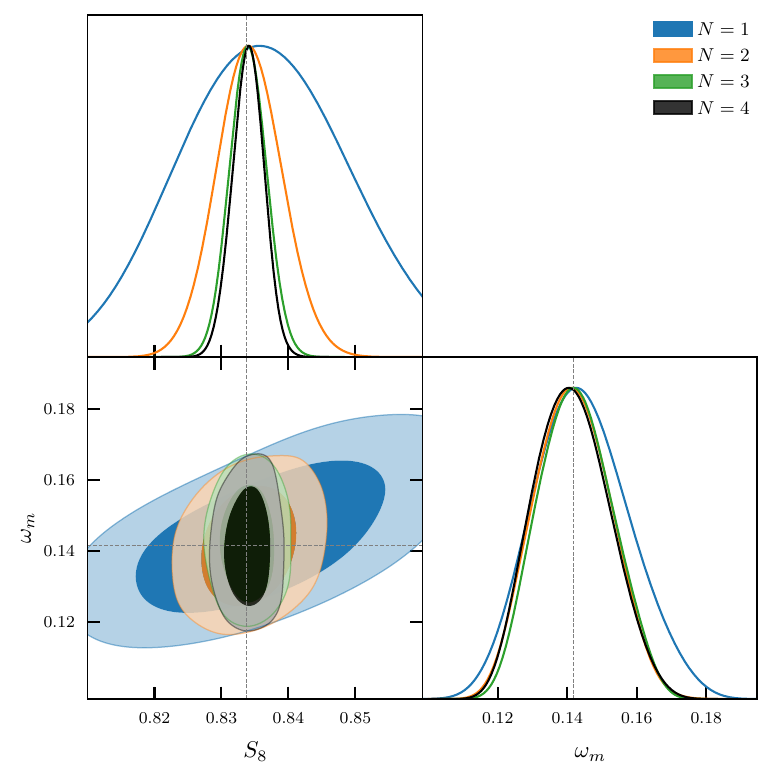}

    \caption{Dependence of $S_8$ and $\omega_m$ constraints without baryon feedback on the lensing counterterm order, where $N=1$ is an analysis with no lensing counterterms. The left and right hand side figures are DES-Y3 and LSST-Y10-like analyses respectively. In the DES case, the $N=4$ expansion yields $25\%$ smaller $S_8$ uncertainties compared to the no-counterterm $N=1$ case, while the $S_8$ error is $82\%$ smaller than the $N=1$ case for LSST-Y10.}
    \label{fig:lct_order_sens}
\end{figure*}

Figure~\ref{fig:lct_order_sens} shows the results of these investigations, with DES-like analyses shown on the left and LSST-like analyses shown on the right. In both cases, the general trend is that the constraining power improves significantly as we increase the order of the LCT expansion. In the DES case, the $N=4$ expansion yields $25\%$ smaller $S_8$ uncertainties compared to the no-counterterm $N=1$ case, while the $S_8$ error is $82\%$ smaller than the $N=1$ case for LSST-Y10. The difference in the two cases is largely driven by the fact that our scale cut criteria are more stringent for the LSST case. 

While this may be counter-intuitive, since we are adding progressively more nuisance parameters as we increase the LCT expansion, this behavior can be explained by observing that the fraction of signal coming from $k>\Lambda$ for a given $\ell$ does not approach 1 until significantly after $C^{ij}_{\ell,\rm UV}$ exceeds 5\%, as shown in Figure~\ref{fig:lensing_ct_k1p0}. The upshot of this is that there is a significant amount of $k<\Lambda$ information at scales that are removed from our analysis when imposing scale cuts such that $C^{ij}_{\ell,\rm UV}/C^{ij}_{\ell}< 0.05$, as we do when we do not employ our LCT expansion. In addition to this observation, in order for increasing $N$ to yield improved cosmological constraining power, it must also be the case that the LCTs are minimally degenerate with the cosmological parameters of interest. If this was not the case and $C^{ij}_{\ell,\rm UV}$ had a very similar shape to the $k<\Lambda$ contributions, then even if we included all scales in our analysis we would not gain additional information about $k<\Lambda$, since this would be indistinguishable from contributions to $C^{ij}_{\ell,\rm UV}$. Throughout this work, we assume $N=4$ when marginalizing over LCTs.

\begin{figure*}
    \centering
    \includegraphics[width=0.45\linewidth]{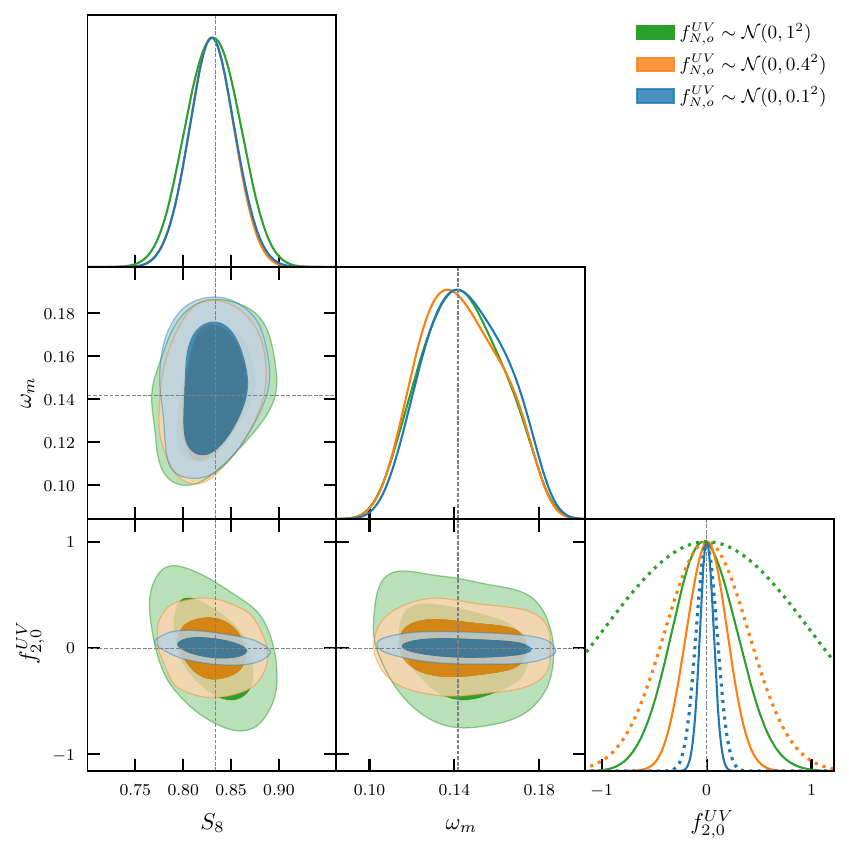}
    \includegraphics[width=0.45\linewidth]{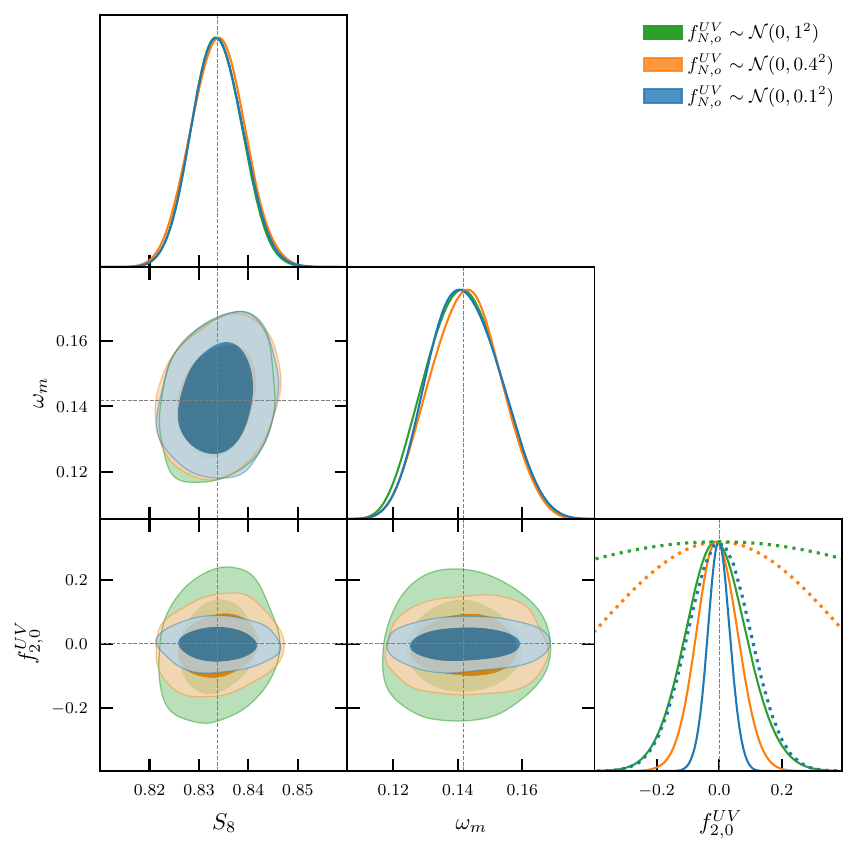}

    \caption{Dependence of $S_8$ and $\omega_m$ constraints on the priors assumed for the lensing counterterm contributions, showing priors with standard deviations of 10\% (blue), 40\% (orange) and 100\% (green) of the fiducial dark matter only signal illustrated as the dotted lines in the posterior of the leading order counterterm, $f^{UV}_{2,0}$. The left and right hand side figures are DES-Y3 and LSST-Y10-like analyses respectively. In the DES case, there is a slight change in constraining power as a function of the LCT priors, with the 100\% prior showing approximately a $18\%$ degradation in constraining power over the other two cases. In the LSST case, all priors yield very nearly the same constraints on the cosmological parameters, as the data can constrain the LCTs at the $\sim 20\%$ level on its own, so setting tight priors on these parameters yields little gain.}
    \label{fig:lct_prior_sens}
\end{figure*}

We also show that our fiducial priors on the LCTs are not informative, i.e. these priors do not drive our conclusion that increasing the LCT order improves constraining power. In order to show this, we vary the widths of our priors on $f^{\rm UV}_{N,o}$, studying cases where $\sigma[f^{\rm UV}_{N,o}] \in \{0.1, 0.4, 1.0\}$, where $\sigma[f^{\rm UV}_{N,o}]=0.4$ is our fiducial choice, again taking $\Lambda=1\himpc$. Figure~\ref{fig:lct_prior_sens} shows the results of these investigations, showing that our $S_8$ constraints are largely insensitive to the widths of our LCT priors. This is due to a combination of the fact that the LCTs are largely not degenerate with the cosmological parameters of interest, and because the data constrain the LCTs to be significantly smaller than the prior widths in all cases except our DES-like analysis with $10\%$ priors, as can be seen by comparing the solid 1D posteriors for the leading order counterterm contribution, $f^{\rm UV}_{2,0}$, to the priors shown by dashed lines on the same panels. In the DES case, the 100$\%$ prior leads to a degradation of constraining power of $18\%$, and so for this less constraining analysis we are somewhat sensitive to our prior, but for LSST-like analyses this sensitivity is entirely removed.

\begin{figure*}
    \centering
    \includegraphics[width=0.45\linewidth]{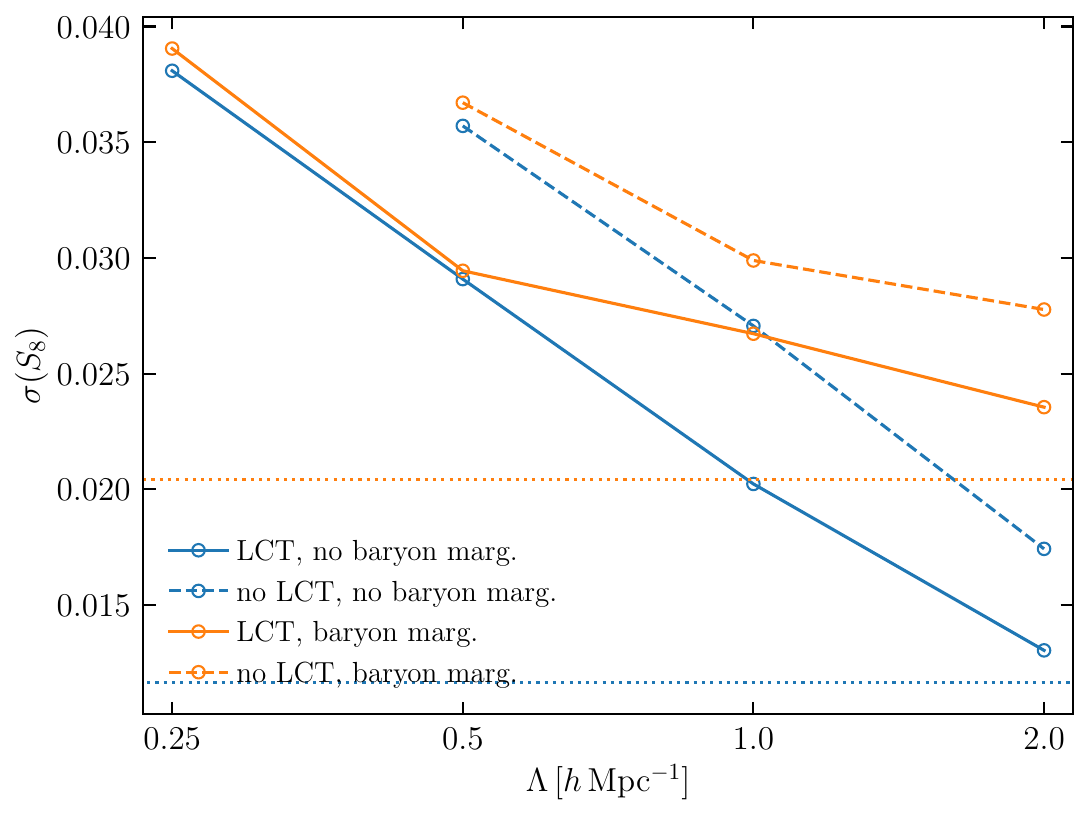}
    \includegraphics[width=0.45\linewidth]{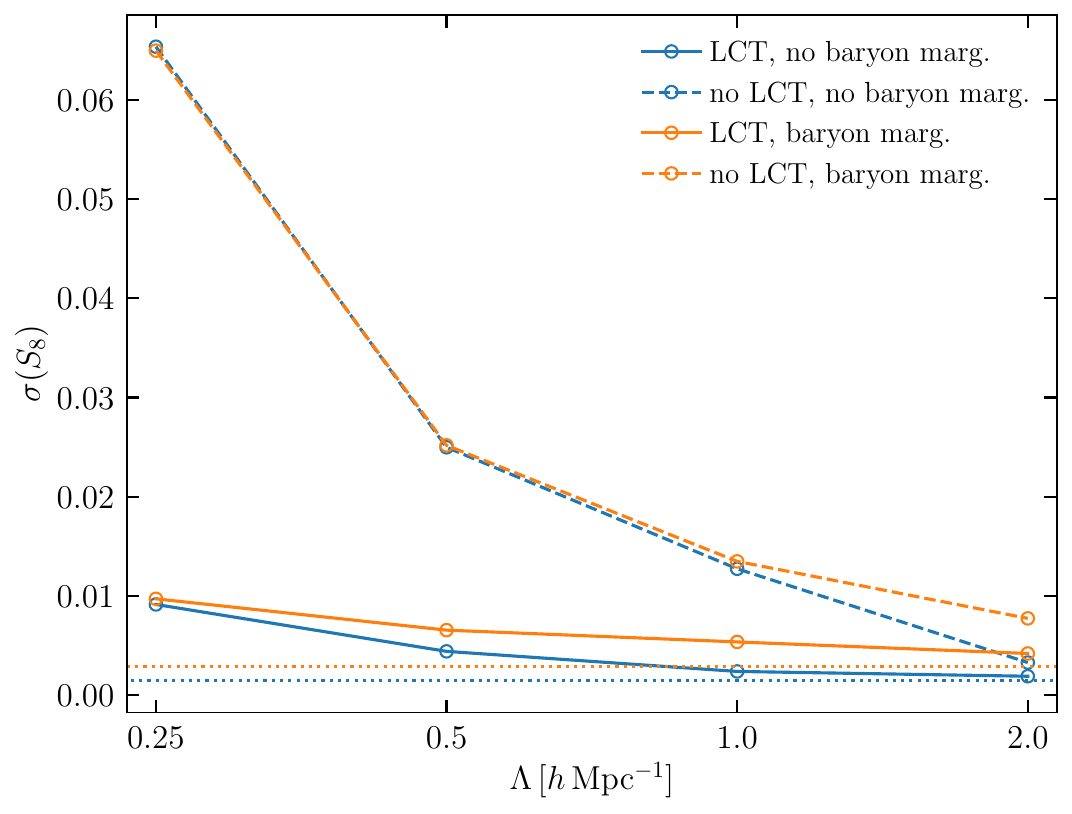}
    \caption{Errors on $S_8$ as a function of the cutoff scale in our model, $\Lambda$, with (solid) and without (dashed) lensing counterterms. Blue lines represent analyses that do not marginalize over baryons, while orange use our fiducial Pad\'e approximant baryon counterterm model. The horizontal dotted lines show the $S_8$ error when fitting to $\ell=2500$ with no lensing counterterms. The left panel shows constraints for DES-Y3-like analyses, while the right shows the same thing for a LSST-Y10-like setup. Dotted lines are not plotted for all values of $\Lambda$, because when we do not include LCTs, there are no data points remaining after scale cuts using our fiducial $\ell$ binning for $\Lambda<0.5\, \himpc$ for DES. The corresponding numerical values are collected in Table~\ref{tab:forecast_summary}.}
    \label{fig:kmax_sens}
\end{figure*}

We also investigate how the improvement in constraining power shown in Figure~\ref{fig:lct_order_sens} depends on the value assumed for $\Lambda$, in addition to whether or not we marginalize over nuisance parameters related to baryonic impacts on the matter power spectrum. These investigations are summarized in Figure~\ref{fig:kmax_sens}, where we again show results for a DES-Y3-like analysis on the left and an LSST-Y10-like analysis on the right. The $x$-axis shows the cutoff scale assumed, and the $y$-axis shows the error on $S_8$ for each analysis. Solid lines use LCTs while dashed lines show the same analysis set up but with no LCTs. Dotted horizontal lines show the constraining power when not marginalizing over LCTs and taking $\ell_{\rm max}=2500$ for all bins as a proxy for the maximum information content available. Analyses that do not marginalize over baryonic impacts, as assumed in the previous figures, are shown in blue. In the LSST case we see that as we increase $\Lambda$ the improvement in constraining power that we obtain from using LCTs decreases, as one might expect given that the LCTs are marginalizing over a smaller contribution at fixed $\ell$ as $\Lambda$ increases, and pushing to higher $\ell$ yields marginally less information due to shape noise contributions to the statistical errors. For DES-Y3 the LCT improvement stays roughly constant across the range of $\Lambda$ that we consider, in part because scale cuts leave no data at the smallest $\Lambda$ for the no-LCT analysis. Another thing to note is that for DES and LSST, there are values of $\Lambda$ which effectively leave no data within our scale cuts for the analyses without LCTs, and so these points are not shown for these cases.

The orange lines in Figure~\ref{fig:kmax_sens} show a more realistic scenario where we marginalize over our fiducial baryon feedback parameterization. In this case, we observe that our constraining power is significantly degraded for $\Lambda\geq0.5\, \himpc$. The difference between the $\ell_{\rm max}=2500$, $N=1$ LSST analysis represented by the dotted orange line and the $\Lambda=2\himpc$ analysis with baryons and LCTs is driven by the ability of very high $\ell$ data to break the degeneracy between $S_8$ and baryons. The $\ell_{\rm max}=2500$ analysis uses scales $k>2\himpc$ where we do not expect this baryon model to hold. The degeneracy between a more complex baryon model and $S_8$ would likely not be broken by including these scales, likely leading to asymptotic $S_8$ constraints closer to those found with $\Lambda=2\himpc$.

\label{subsec:noiseless_tests}

\begin{table*}[htb!]
    \centering
    \begin{tabular}{|c|cc|cc|cc|cc|}
        \hline
        \hline
         & \multicolumn{4}{c|}{LSST-Y10} & \multicolumn{4}{c|}{DES-Y3} \\
         & \multicolumn{2}{c|}{no baryon marg.} & \multicolumn{2}{c|}{w/ baryon marg.} & \multicolumn{2}{c|}{no baryon marg.} & \multicolumn{2}{c|}{w/ baryon marg.} \\
         $\Lambda$ [$h\,\mathrm{Mpc}^{-1}$] & no LCT & $N{=}4$ & no LCT & $N{=}4$ & no LCT & $N{=}4$ & no LCT & $N{=}4$ \\
         \hline
         0.25 & 0.0654 & 0.0091 ($86\%$) & 0.0650 & 0.0097 ($85\%$) & --     & 0.0381             & --     & 0.0390 \\
         0.5  & 0.0250 & 0.0044 ($82\%$) & 0.0252 & 0.0065 ($74\%$) & 0.0357 & 0.0291 ($19\%$)  & 0.0367 & 0.0294 ($20\%$) \\
         1.0  & 0.0127 & 0.0023 ($82\%$) & 0.0135 & 0.0053 ($60\%$) & 0.0271 & 0.0202 ($25\%$)  & 0.0299 & 0.0267 ($11\%$) \\
         2.0  & 0.0032 & 0.0019 ($43\%$) & 0.0077 & 0.0042 ($46\%$) & 0.0174 & 0.0130 ($25\%$)  & 0.0278 & 0.0235 ($15\%$) \\
         \hline
    \end{tabular}
    \caption{Marginalized $\sigma(S_8)$ from noiseless-mock LSST-Y10 and DES-Y3 cosmic shear analyses as a function of the model cutoff $\Lambda$, with and without lensing counterterms ($N{=}4$) and with and without marginalization over the fiducial Pad\'e baryon counterterm model. Parenthesized percentages are the fractional improvement in $\sigma(S_8)$ over the corresponding no-LCT analysis at the same $\Lambda$ and baryon treatment. Dashes indicate configurations for which our fiducial scale-cut criteria leave no data (see Figure~\ref{fig:kmax_sens}). At our fiducial $\Lambda=1\,h\,\mathrm{Mpc}^{-1}$ with no baryon marginalization, LCTs shrink $\sigma(S_8)$ by $82\%$ for LSST-Y10 and $25\%$ for DES-Y3.}
    \label{tab:forecast_summary}
\end{table*}

\subsection{Contamination from Baryons and Non-standard Physics}
Having demonstrated the ability of our LCT methodology to extract more constraining power from $k<\Lambda$ contributions to our data vectors, we now demonstrate the ability of LCTs to marginalize over potential biases from $k>\Lambda$, while still preserving information about these contributions. For these tests, we use the redshift distributions and covariance matrix for the LSST-Y10 set up described in the previous section. We first consider biases to the matter power spectrum due to baryonic physics not included in our dark--matter--only matter power spectrum model, measured from hydrodynamical simulations. In this subsection only, we fix the scale cuts to those appropriate for $N=4$, even when fitting without LCTs in order to demonstrate the ability of our LCTs to marginalize over systematic biases that would otherwise be large without their inclusion.

\begin{figure}
    \centering
    \includegraphics[width=\linewidth]{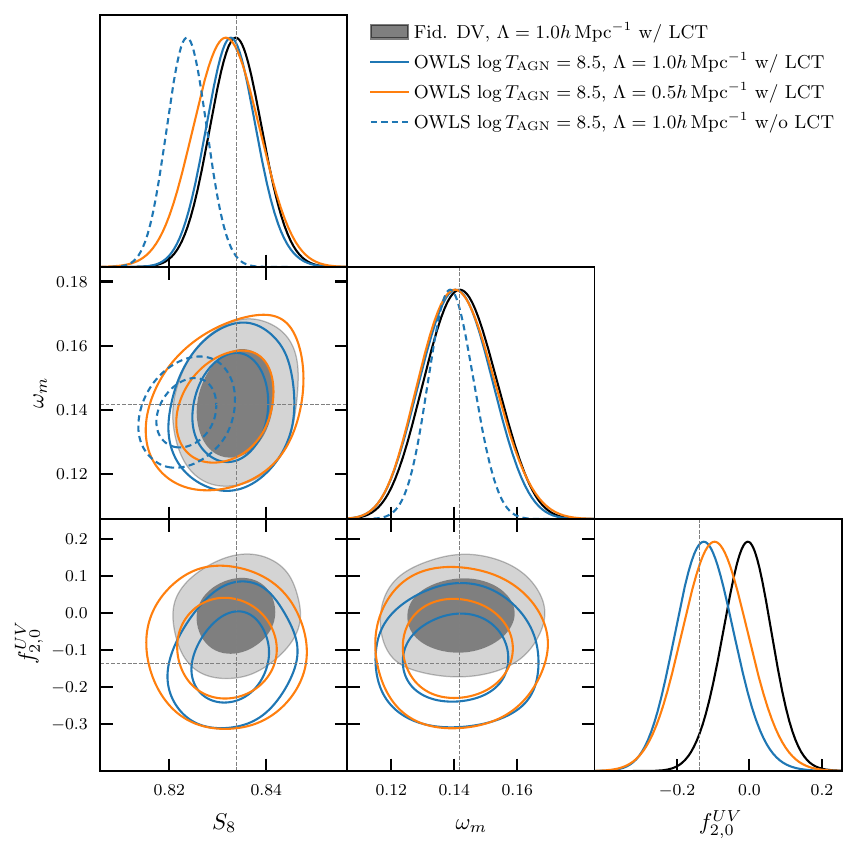}
    \caption{Comparison of constraints from an LSST-Y10-like cosmic shear analysis fitting to a data vector with a power spectrum suppression measured from the OWLS AGN $\log T_{\rm AGN}=8.5$ \cite{LeBrun2014} simulation with (solid) and without (dashed) lensing counterterms, compared to fiducial constraints with lensing counterterms fitting to a dark--matter--only data vector (black). The $k>0.5\, \himpc$ fits (orange) use our polynomial baryon marginalization model, which deviates from the OWLS AGN suppression at $k>0.5\, \himpc$ more significantly than the Pad\'e expansion, used for the $k>1\, \himpc$ constraints (blue), such that the $k>0.5$ no LCT chain did not converge. See text for a more detailed explanation of this phenomenon.}
    \label{fig:owls_lsst}
\end{figure}

For this example, we measure the suppression of the matter power spectrum in the $\log T_{\rm AGN}=8.5$ OWLS AGN simulation \cite{LeBrun2014}, with respect to the paired dark--matter-only version of this simulation:
\begin{equation}
    S(k,z) = \frac{P_{\rm mm,\, hydro}(k,z)}{P_{\rm cdm}(k,z)}\, 
\end{equation}
and produce a contaminated data vector with our model, taking $P_{\rm mm}(k) = S(k,z)P_{\rm cdm}(k)$. The results of fitting this contaminated data vector are shown in Figure~\ref{fig:owls_lsst}, where we perform analyses with our fiducial model taking $\Lambda=0.5\, \himpc$ in orange and $\Lambda=1.0\, \himpc$ in blue. Solid and dashed lines represent analyses with and without LCTs respectively. The grey solid contour is an analysis of an uncontaminated data vector with $\Lambda=1.0\, \himpc$ and including LCTs for reference. For the $\Lambda=0.5\, \himpc$ case, we use our polynomial baryon counterterm model, whereas for  $\Lambda=1.0\, \himpc$ we use our fiducial Pad\'e baryon expansion.

We see that for $\Lambda=1.0\, \himpc$ the LCT case is unbiased while without LCTs we see a large bias. This result is further exacerbated when using $\Lambda=0.5\, \himpc$ where we use our polynomial baryon counterterm model. The no LCT case is so biased that our chain did not converge in a reasonable amount of time, while the analysis with LCTs remains unbiased. This behavior is a result of the polynomial baryon counterterm model, which is a good fit to baryonic suppressions at $k<0.5\, \himpc$, but for $k>0.5\, \himpc$ the modifications that are predicted rapidly diverge from the OWLS AGN model. Because we are fitting to the same angular scales with and without LCTs in this analysis, unlike in the rest of the paper, the no LCT analysis is significantly biased by the inability of our polynomial counterterm model to fit the OWLS data vector at $k>0.5\, \himpc$. In the $\Lambda=1.0\, \himpc$ case, the fact that the no LCT analysis is less biased than the $\Lambda=0.5\, \himpc$ case is a result of the less divergent behavior of the Pad\'e approximant counterterm as shown in Figure~\ref{fig:baryon_models} where we compare our counterterm models to ${\rm SP}(k)$.

As a more extreme toy example, we also produce contaminated data vectors with an approximation to the impact of axions on the matter power spectrum following \cite{Passaglia2022}:
\begin{equation}
    S(k,z) = \frac{\sin(x^n)}{x^n(1+Bx^{6-n})}\,
\end{equation}
\noindent where 
\begin{equation}
    x = A\frac{k}{k_J},\qquad  k_{J} = 9m^{1/2}_{22},\,
\end{equation}
\noindent and 
\begin{align}
    A &= 2.22 m_{22}^{1/25-1/1000\ln(m_{22})}\\ \nonumber
    B &= 0.16m_{22}^{-1/20},
\end{align}
with $m_{22} = m/10^{-22}\rm eV$, and the power-law index is $n=5/2$. For these tests we take m=$2.5\times10^{-23}\rm eV$, such that the power spectrum suppression for $k<1\himpc$ is negligible, and then turns on rapidly, growing to $>20\%$ at $k=2\himpc$. We caution that this should be taken only as an extreme test case, since we do not properly take into account the effect of nonlinearities (see e.g. ref.~\cite{Marsh16} for a treatment within the halo model).

\begin{figure}
    \centering
    \includegraphics[width=\linewidth]{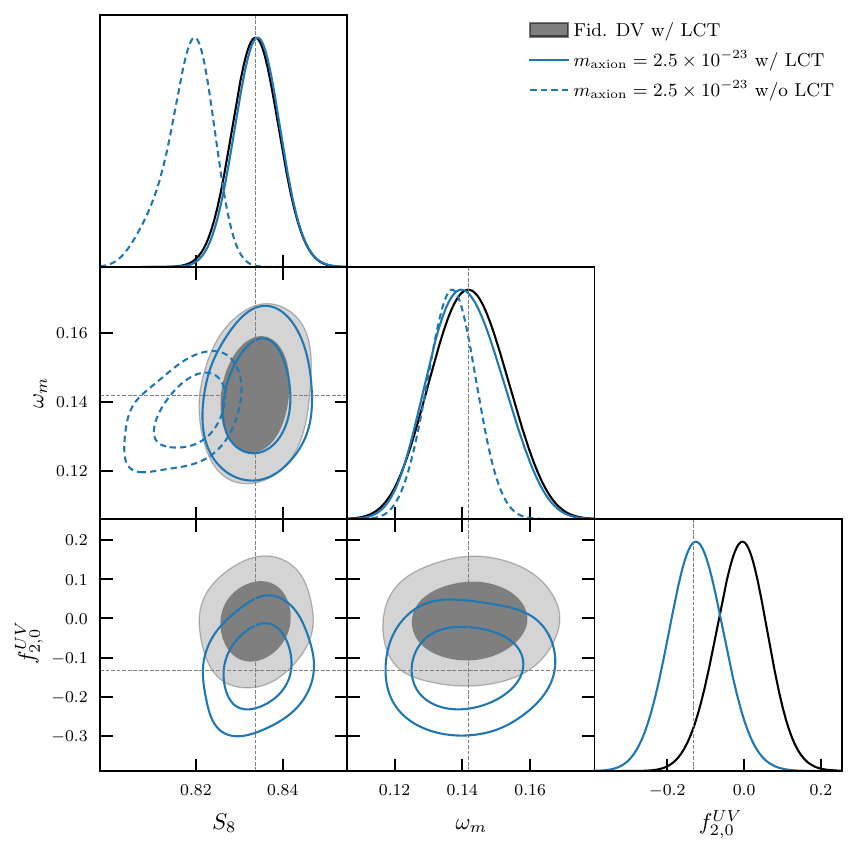}
    \caption{Comparison of constraints from an LSST-Y10-like cosmic shear analysis fitting to a data vector mimicking a $2.5\times10^{-23}\, \rm eV$ axion with (solid blue) and without (dashed blue) lensing counterterms, compared to fiducial constraints with lensing counterterms fitting to a $\Lambda$CDM data vector (black). When including lensing counterterms, our $\Lambda$CDM constraints are unbiased and the counterterm posteriors agree with the expected value given the assumed axion model (crosshair).}
    \label{fig:axion_lsst}
\end{figure}

The results of fitting an LSST-Y10-like data vector using $\Lambda=1\himpc$ with this contamination are shown in Figure~\ref{fig:axion_lsst}, where solid blue contours represent an analysis with LCTs and dashed show the analogous analysis without. The grey filled contours show an analysis on a $\Lambda$CDM data vector with LCTs. As in the previous tests, we fix the intrinsic alignment contributions to zero, but otherwise use our fiducial model. We again fit to the same angular scales with and without LCTs, unlike in the rest of this work where we derive different scale cuts depending on whether we include LCTs or not. Here we see that the no LCT case is significantly biased, the data vector analyzed here contains significant contributions from $k>\Lambda$ in our scale cuts, and the axion suppression is not well fit by our Pad\'e baryon counterterm model, unlike in the OWLS AGN contamination case.

Notably, we see significant shifts in the posterior of the first order lensing counterterm, $f_{2,0}^{\rm UV}$, such that the measured posterior distribution is peaked at the value of this parameter that is expected from the axion suppression that we have introduced. Thus, despite the fact that our LCT model marginalizes over $k>\Lambda$ contributions to our data, we can still potentially extract information from these scales through the constraints on the LCTs themselves. In this case the interesting $k>\Lambda$ contributions are sourced solely by our axion approximation, but in an actual data analysis one would need to fold in baryonic suppression in order to correctly interpret the LCT constraints. Nevertheless, the fact that we can place interesting constraints on the LCTs, and they can be interpreted in terms of beyond dark--matter--only $\Lambda$CDM physics, opens the door for more ambitious modeling efforts to extract interesting physical constraints from $k>\Lambda$. We also note that in Figure~\ref{fig:owls_lsst}, the $f_{2,0}^{\rm UV}$ posterior is consistent with our expectation given the OWLS AGN model, as indicated by the dotted vertical line, where in this case we show the expectation for $\Lambda=1\himpc$, with the $\Lambda=0.5\himpc$ value being somewhat smaller. Thus, posteriors on these counterterms can also be used to gain insight into baryonic effects on the matter power spectrum.

\subsection{Fitting $N$-body Simulations}
\label{sec:model_val}
In addition to fitting noiseless data generated by our model, we wish to demonstrate the accuracy of the various approximations made in our modeling by fitting to realistic simulated data. To do this, we measure cosmic shear angular power spectra from the \texttt{Buzzard v2.0} simulations, using the DES-Y3 like source catalogs described in \cite{DeRose2022}. We apply the same measurement pipeline to these simulations as that used on the DES-Y3 data described in Section~\ref{sec:data}, except here we set all responses and weights equal to one and use the true gravitational shear values from the simulation in order to remove shape noise. We use the true redshift distributions for the sources in our model predictions. For our fits, we average over 10 simulated DES-Y3 footprints, taking the covariance to be the same as that estimated for the DES-Y3 data.

\begin{figure}
    \includegraphics[width=\linewidth]{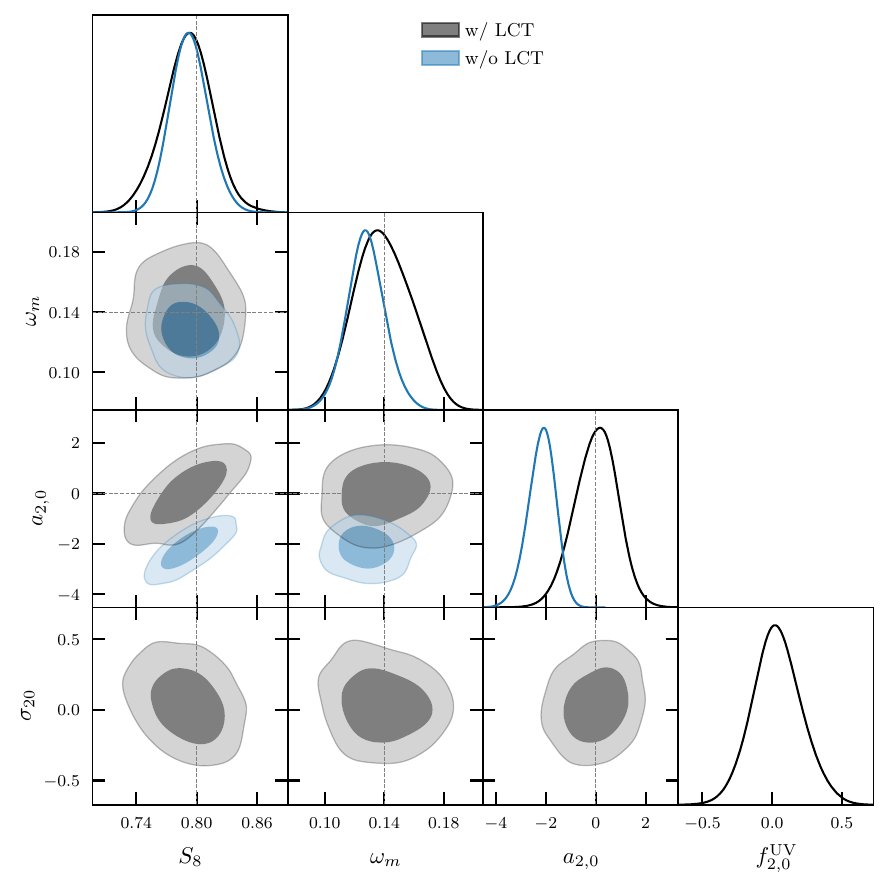}
    \caption{Constraints on the Buzzard simulations, truncating contributions to the limber integrals above $\Lambda$ entirely, rather than our fiducial treatment. In this case, including counterterms is absolutely required in order to obtain unbiased constraints.}
    \label{fig:buzzard_val}
\end{figure}

We fit these data using $\Lambda=1\, h\, \rm Mpc^{-1}$, setting contributions from $k>\Lambda$ to zero and shifting our $f_{N,o}^{\rm UV}$ priors to have means of one instead of zero. This is distinct from the rest of the fits in this work, where we include $k>\Lambda$ contributions from our fiducial model with baryon counterterms set to zero, as described in Section~\ref{sec:params}. We make this choice here in order to demonstrate that the lensing counterterms are indeed doing their intended job of marginalizing over $k>\Lambda$ contributions to our measurements. The results of these fits are shown in Figure~\ref{fig:buzzard_val}, with $N=4$ results shown in black, and no lensing counterterms results ($N=1$) shown in blue, otherwise using our fiducial model with intrinsic alignments set to zero.  We see that even when completely truncating model contributions from $k>\Lambda$, when including lensing counterterms with $N=4$ we are able to recover the true cosmology of the \texttt{Buzzard v2.0} simulations, with the LCTs all centered around the expected value of one and baryon counterterms all centered around zero. On the other hand, when we do not include LCTs and otherwise keep all modeling choices identical we see a $1\sigma$ bias in $\omega_m$. Given the significantly reduced noise on these simulations, given that there are 10 realizations without shape noise, this constitutes a strongly detected systematic bias when not employing LCTs. Our baryon parameterization is able to absorb what would have been a bias in $S_8$, as can be seen by the fact that $a_{2,0}$ is many sigma away from zero for the no LCT case, though we emphasize that this is a result of improperly blending physics above and below $\Lambda$, in this case through greatly overestimating the impact of baryonic feedback on large scales. This demonstrates that our LCT contributions are correctly accounting for the $k>\Lambda$ signal in our simulated measurements, and that our dark matter only model is accurate enough to recover the true cosmology of the \texttt{Buzzard v2.0} simulations.

\section{Constraints on DES-Y3 Data}
\label{sec:data_results}
\begin{figure}
    \centering
    \includegraphics[width=\linewidth]{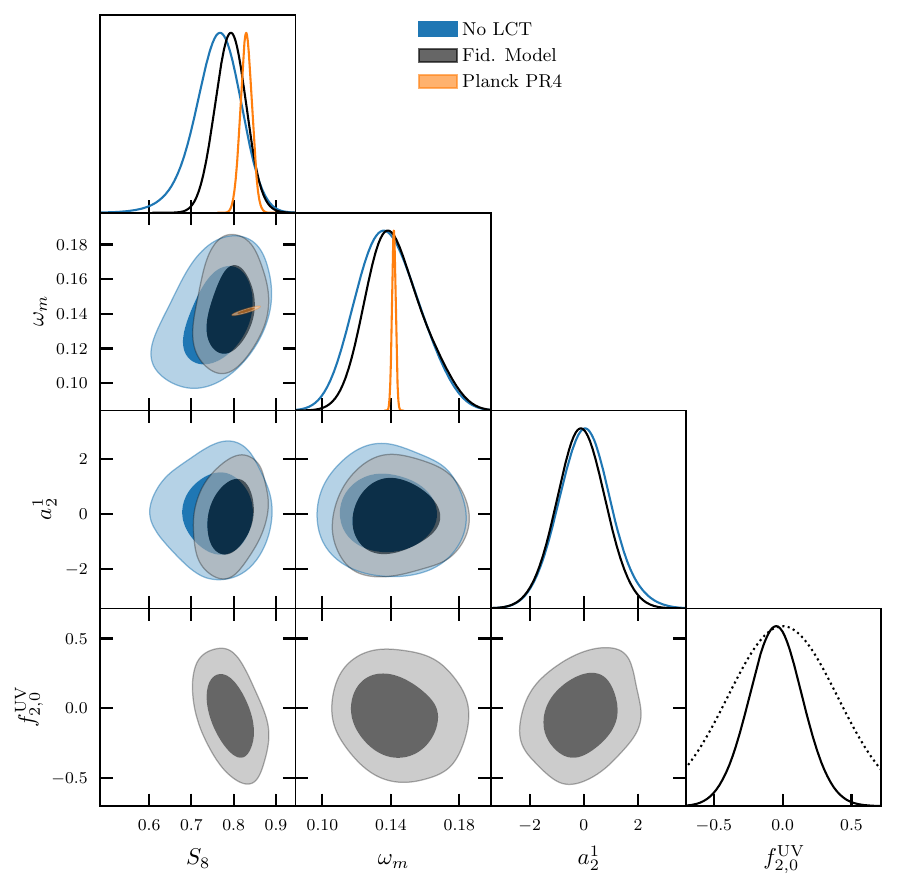}
    \caption{Fitting DES-Y3 cosmic shear power spectra with (black) and without (blue) lensing counterterms, compared to Planck PR4 $\Lambda$CDM constraints \cite{Tristram2023}. When including lensing counterterms, we see an improvement in constraining power of $19\%$ in $S_8$. Both sets of constraints agree with Planck at the $1\sigma$ level, although the constraints with lensing counterterms prefer a $4\%$ higher value of $S_8$, which is more consistent with Planck. The first order lensing counterterm contribution $f^{\rm UV}_{2,0}$ is consistent with $\Lambda$CDM ($f^{\rm UV}_{2,0}=0$) and is constrained significantly better than the prior, represented by the black dotted line on the marginalized one-dimensional posterior of $f^{\rm UV}_{2,0}$.}
    \label{fig:desy3_lct_comp}
\end{figure}

We now turn to a demonstration of our LCT model on the public DES-Y3 data. Figure~\ref{fig:desy3_lct_comp} shows a comparison of the constraints on our DES-Y3 measurements, described in Section~\ref{sec:data}, with and without lensing counterterms assuming $\Lambda=1\himpc$. As in our simulated analyses, our scale cuts vary depending on whether we do or do not include LCTs as described in Section~\ref{sec:scale_cuts}. When using LCTs in this section, we always take $N=4$ for reasons described in Section~\ref{sec:counterterms}. 

Our fiducial analysis gives 
\begin{align*}
    S_8 = 0.798\pm 0.026\, (0.826) \\
    \omega_m = 0.143^{+0.015}_{-0.018}\, (0.133)
\end{align*}
compared to
\begin{align*}
    S_8 = 0.780^{+0.035}_{-0.030}\, (0.797) \\
    \omega_m = 0.138^{+0.017}_{-0.019}\, (0.137)
\end{align*}
\noindent without LCTs. Figure~\ref{fig:desy3_lct_comp} also shows that the first order lensing counterterm contribution $f_{2,0}^{\rm UV}$ is consistent with our fiducial $\Lambda$CDM model prediction ($f_{2,0}^{\rm UV}=0$) and is constrained significantly better than the prior on this parameter, represented by the dashed black line in the one-dimensional posterior distribution. Both sets of constraints are consistent with Planck PR4 \cite{Tristram2023}, with our constraints using LCTs preferring a $2.3\%$ larger mean $S_8$ than without LCTs. When including LCTs our constraints on $S_8$ and $\omega_m$ improve by a factor of $19\%$ and $12\%$ respectively, while the constraint on the best-constrained baryon counterterm spline node $a_{2}^{1}$ improves by $7\%$.

This improvement is larger than the improvements found on noiseless simulations in Figure~\ref{fig:kmax_sens} when marginalizing over our fiducial baryon counterterm model, largely because in this section we also marginalize over our fiducial intrinsic alignment model whereas in the noiseless simulated data we did not. A significant amount of the improvement in constraining power seen here is thus attributable to improved constraints on intrinsic alignment parameters. Although the IA contributions to our signals are generically sensitive to lower $k$ than the lensing contributions without LCTs, when including LCTs this is no longer the case. While none of the intrinsic alignment parameters are detected at the $0.5\sigma$ level, this is cause for some concern given the perturbative nature of our intrinsic alignment model. For now we simply note this fact, but we will return to it in a future work that is focused on a full accounting of IA uncertainties in cosmic shear. We also note that for future datasets such as LSST, where the LCT methodology presented here will be even more powerful as shown in Figure~\ref{fig:lct_order_sens}, we expect source redshift distributions that are significantly better localized, making the increased sensitivity to high-$k$ contributions in the IA model less of a concern.

\begin{figure*}
    \includegraphics[width=\linewidth]{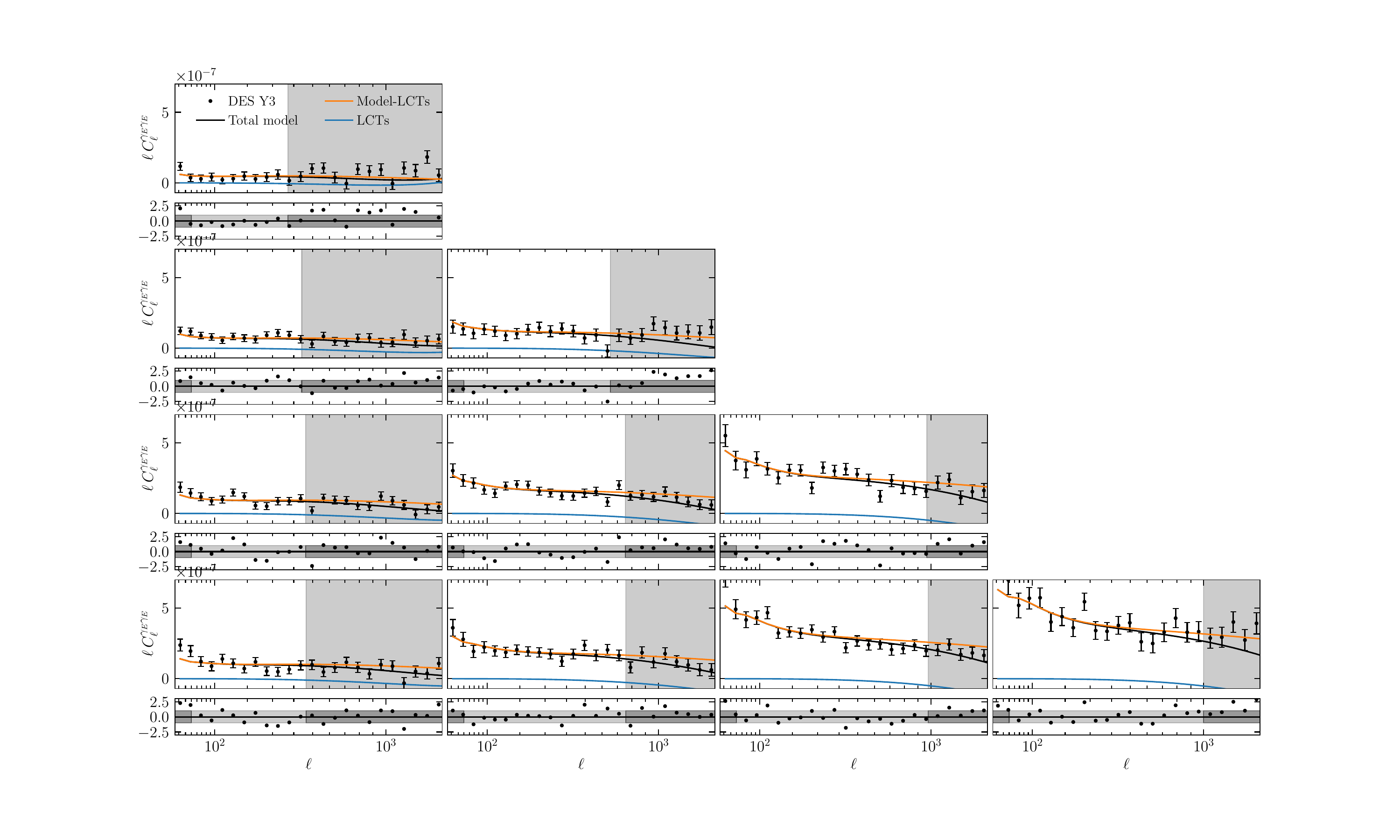}
    \caption{Best fit fiducial model to the DES-Y3 data for $\Lambda=1\, \himpc$. Data points falling in grey regions are not included in our fits. The best fit has $\chi^2=136.8$ for $130$ data points with $n_{\rm par, eff}=11.0$ yielding $\chi^2_{\rm red} = \chi^2/(n_{\rm dv}-n_{\rm par,eff}-1)=1.15$ ($p=0.125$). Black lines represent our total model prediction, while blue and orange lines show the model without the lensing counterterms and lensing counterterm contribution respectively.}
    \label{fig:best_fit}
\end{figure*}

Figure~\ref{fig:best_fit} shows our best fit fiducial model to our DES-Y3 measurements with each panel showing a different redshift bin combination. The best fit model yields $\chi^2=136.8$ for $130$ data points included in the fit. The total number of model parameters is 35, but many of these are not constrained appreciably. In order to account for this, we calculate the number of effective parameters as 
\begin{equation}
    n_{\rm par, eff} = n_{\rm par} - \rm Tr(\Sigma_{\rm prior}^{-1}\Sigma_{\rm post})\,
\end{equation}
\noindent where $n_{\rm par}$ is the total number of parameters sampled over, $\Sigma_{\rm prior}$ is the prior covariance matrix, which is diagonal in this case, and $\Sigma_{\rm post}$ is the posterior parameter covariance matrix. This yields $n_{\rm par, eff}=11.0$ for our fiducial model, which gives $\chi^2_{\rm red} = \chi^2/(n_{\rm dv}-n_{\rm par,eff}-1)=1.15$ ($p=0.125$). 

\begin{figure}
    \centering
    \includegraphics[width=\linewidth]{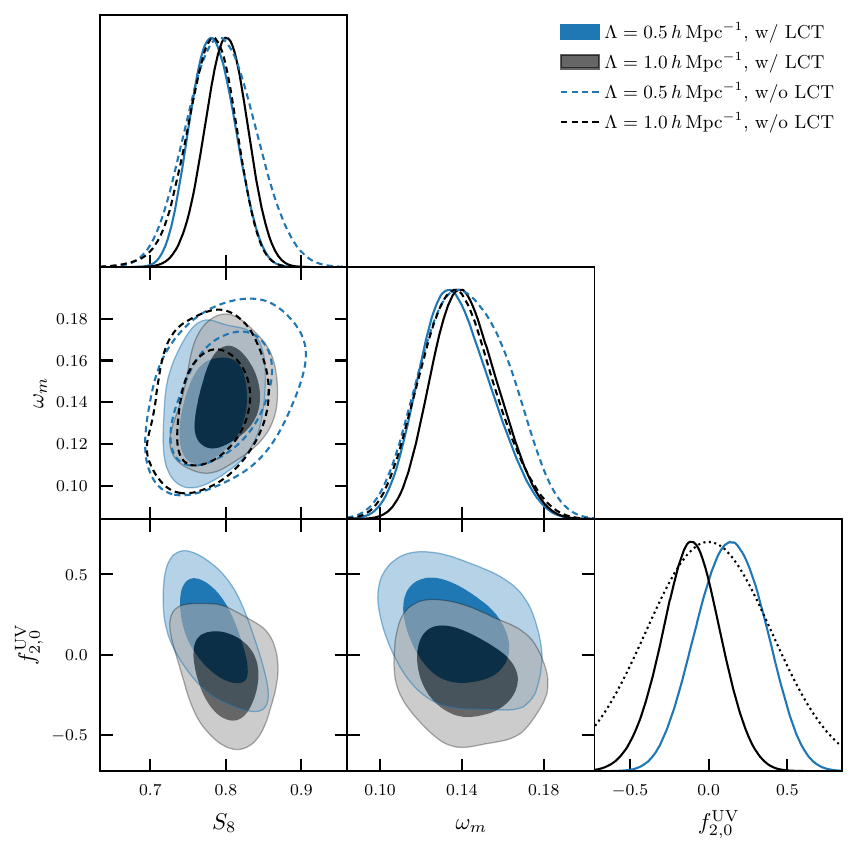}
    \caption{Fitting DES-Y3 setting $\Lambda=1\,\himpc$ (black) and  $\Lambda=0.5\,\himpc$ (blue) with (solid) and without (dashed) lensing counterterms. The $S_8$ constraining power of the $\Lambda=0.5\,\himpc$ analysis with lensing counterterms only degrades by $11\%$ compared to the more aggressive $\Lambda=1.0\,\himpc$ analysis, while the $S_8$ constraining power of the analysis without lensing counterterms degrades by a much more significant $35\%$.}
    \label{fig:desy3_lambda_comp}
\end{figure}

The LCT methodology presented in this work enables the clean separation of scales that can be modeled with relatively small theoretical uncertainties from those that are dominated by astrophysical systematics. One major advantage of this is that the constraining power of analyses with LCTs degrades much less quickly  as we decrease $\Lambda$ than those without LCTs. In order to demonstrate this, we compare constraints setting $\Lambda=1\,\himpc$ and $\Lambda=0.5\,\himpc$ with and without LCTs in Figure~\ref{fig:desy3_lambda_comp}. With LCTs and $\Lambda=0.5\,\himpc$ we find
\begin{align*}
    S_8 = 0.783\pm 0.029\, (0.815) \\
    \omega_m = 0.137^{+0.015}_{-0.019}\, (0.131)
\end{align*}
\noindent and without LCTs we obtain
\begin{align*}
    S_8 = 0.794\pm 0.043\, (0.841) \\
    \omega_m = 0.142\pm 0.020\, (0.147)
\end{align*}
\noindent We see that the constraining power on $S_8$ degrades by $11\%$ with LCTs and $35\%$ without LCTs when switching from $\Lambda=1\, \himpc$ to $\Lambda=0.5\, \himpc$. The fact that the constraining power degrades by so little when switching to $\Lambda=0.5\,\himpc$ is a major advantage of our LCT methodology, and enables potential perturbative modeling of the cosmic shear signal without drastic degradation of constraining power, a topic that we will revisit in the future. 

\begin{figure*}
    \includegraphics[width=\linewidth]{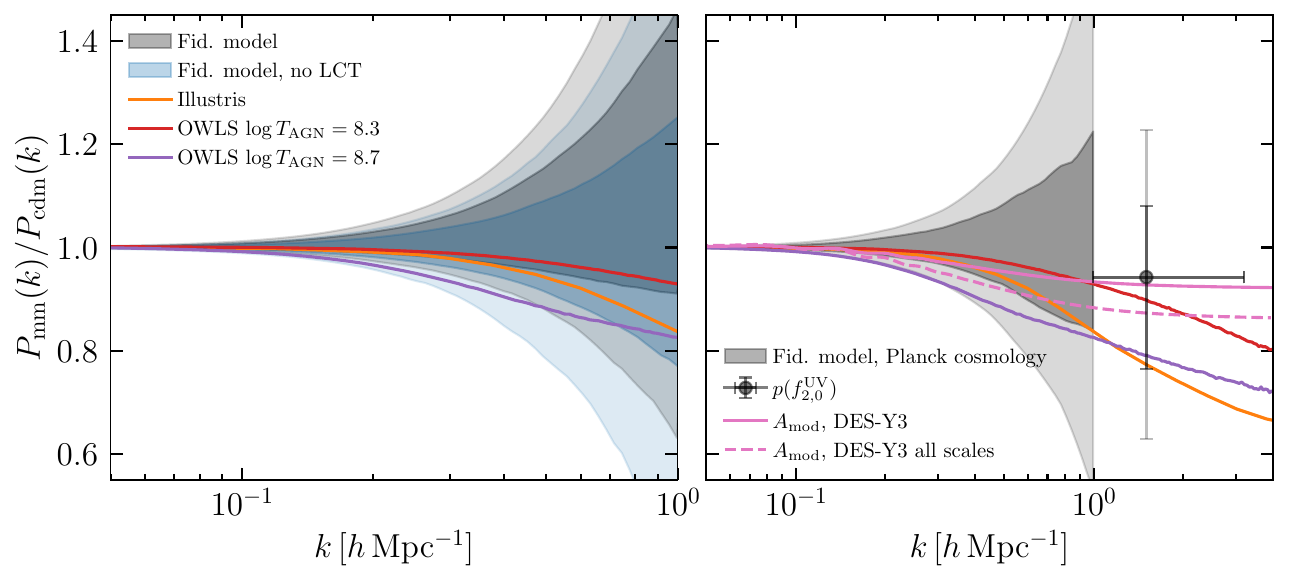}
    \caption{({\it Left}) 1 and 2$\sigma$ (darker and lighter shading) posterior predictive distributions for the ratio of our matter power spectrum predictions at $z=0$ including baryon counterterms to those without. Black envelopes show the constraints using our fiducial model with LCTs and blue show constraints without LCTs both for $\Lambda=1\himpc$. We plot a few predictions from high (Illustris and OWLS $\log T_{\rm AGN}=8.7$) and low (OWLS $\log T_{\rm AGN}=8.3$) feedback hydrodynamical simulations \cite{Pillepich2018,LeBrun2014}. All are consistent with our constraints at the 2$\sigma$ level. {(\it Right}) Same as left but fixing the cosmology to the Planck PR4 best fit \cite{Tristram2023} and including the implied constraint for $k>\Lambda$ from our LCTs as the point at $k\sim1.6\, \himpc$ as described in the text. Here we also plot the ``$A_{\rm mod}$'' constraints from \cite{Preston2023} using the DES-Y3 fiducial scale cuts (solid) and all scales (dashed), which also provide $\sim 1\sigma$ consistent constraints to our model when fixed to the Planck best fit cosmology.}   \label{fig:baryon_ppd}
\end{figure*}

In order to visualize the constraints on deviations from CDM matter power spectrum predictions, we plot the posterior predictive distribution (PPD) of the ratio of our model for $P_{\rm mm}(k,z=0)$ to the CDM predictions, $P_{\rm CDM}(k,z=0)$ in Figure~\ref{fig:baryon_ppd}. The former includes our baryon counterterm model, while the latter does not. The left panel compares posterior predictive distributions for $\Lambda=1\, \himpc$ with (black) and without LCTs (blue). We see that the LCT model provides slightly tighter constraints, consistent with our observation that the posterior on $a_{2}^{1}$ is smaller by a factor of $\sim7\%$ for the LCT analysis, compared to the no LCT analysis. We also compare with predictions from three hydrodynamical simulations, OWLS $\log T_{\rm AGN}=8.3$ and $\log T_{\rm AGN}=8.7$ \cite{LeBrun2014}, as well as the original Illustris simulation, which roughly bracket the range of predicted baryonic feedback strengths. We see that the lensing data does not provide very strong constraints on the shape of the matter power spectrum to rule any of these scenarios out at $>2\sigma$, although we see no evidence of matter power spectrum suppression beyond the CDM prediction (see ref.~\cite{Terasawa25} for a similar conclusion from HSC data) and the OWLS $\log T_{\rm AGN}=8.7$ is just on the lower edge of the $2\sigma$ posteriors.

On the right, we plot the same quantity, but assuming a $\Lambda$CDM cosmological model fixed to the best fit from the Planck PR4 analysis \cite{Tristram2023}. We also plot the implied constraint $(P_{\rm mm}/P_{\rm CDM})(k>\Lambda)$ from our $f_{2,0}^{UV}$, where the extent of the error bars in the x-direction encompasses the $k$-range where the integrand of $\sigma_{2,0}^{\rm UV}$ receives $95\%$ of its contribution. Here too we see no evidence of modifications to the matter power spectrum beyond the CDM prediction. For comparison, we also plot the constraint on $(P_{\rm mm}/P_{\rm CDM})(k)$ from the analysis in \cite{Preston2023} using a phenomenological model where the strength of nonlinear clustering is allowed to deviate from $\Lambda$CDM with a scale-independent amplitude $A_{\rm mod}$ \cite{Amon22}, that uses the fiducial DES-Y3 scale cuts (solid) and all scales (dashed), fixing the cosmology to the best fit from \cite{Efstathiou&Gratton2021}, which is negligibly different from that of \cite{Tristram2023} for the purposes of this analysis. The constraint from  \cite{Preston2023} with scale cuts is consistent with our direct constraint on the small-scale matter power spectrum at the $1\sigma$ level, a difference that is very plausibly explicable by differences in matter power spectrum and intrinsic alignment modeling, while the constraint without scale cuts is marginally less consistent.

\begin{figure*}
    \includegraphics[width=\linewidth]{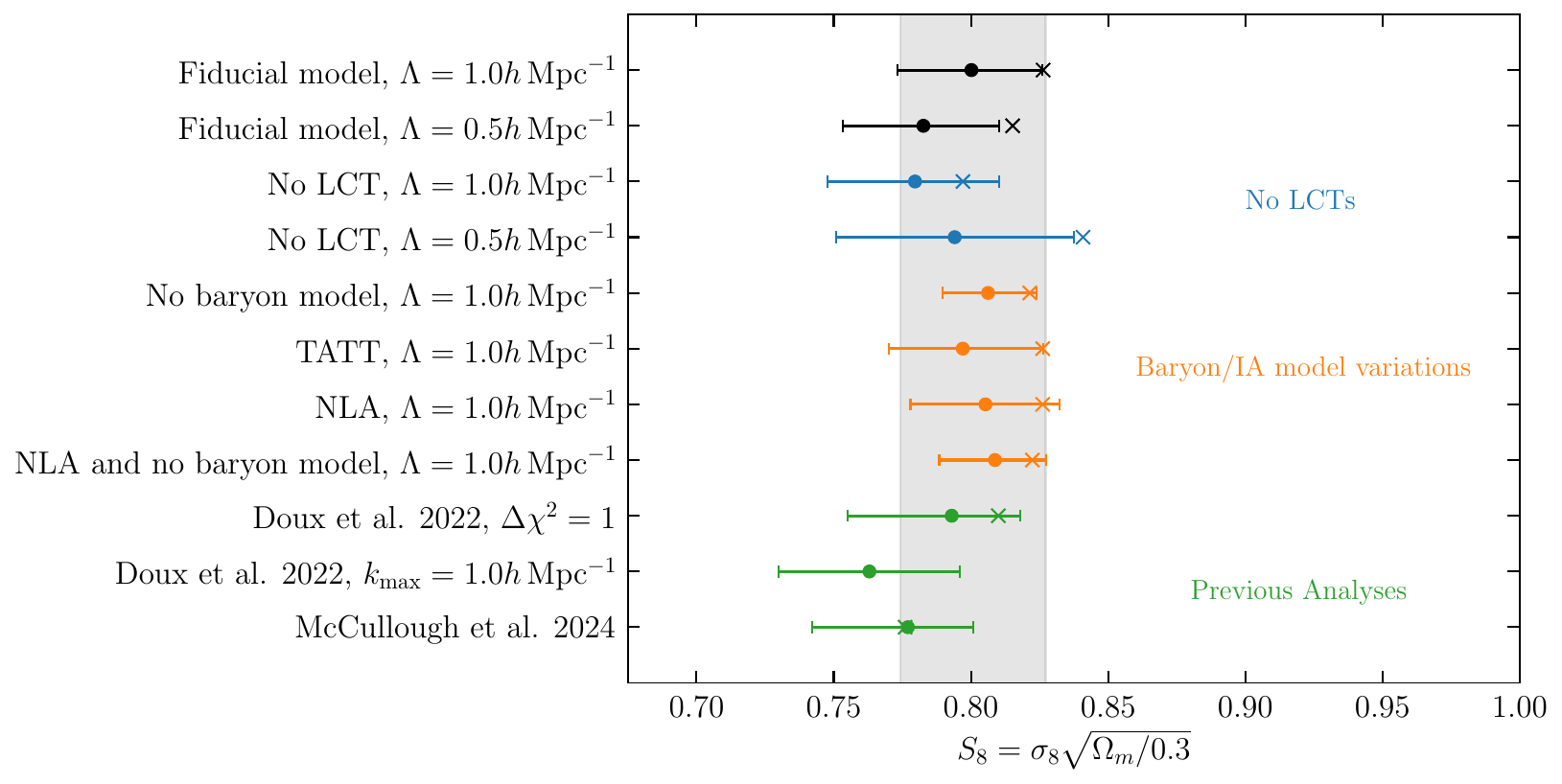}
    \caption{Comparison of $S_8$ constraints with posterior means shown as points, best fit values shown as crosses, and error bars representing $68\%$ confidence intervals. See text for details about the analyses plotted here.}
    \label{fig:desy3_column}
\end{figure*}

Finally, in Figure~\ref{fig:desy3_column} we compare the results from this work to two analyses using the DES-Y3 data, one using similar harmonic space measurements \cite{Doux2021} and another more recent analysis \cite{McCullough2024} that includes a corrected redshift binning from the fiducial DES-Y3 cosmic shear analyses \cite{Amon22} and \cite{Secco22}. Points in this figure represent one-dimensional posterior means, crosses show best fit values obtained by running a LBFGS minimizer, and error bars are $68\%$ confidence intervals. Our fiducial $\Lambda=1\, \himpc$ constraint prefers significantly higher $S_8$ values than the fiducial analyses reported in refs.~\cite{Doux2021,McCullough2024}. This difference appears to be at least somewhat driven by the inclusion of $k>0.5\, \himpc$ data, as our fiducial constraints taking $\Lambda=0.5\, \himpc$ are in better agreement with these previous analyses.

Refs.~\cite{Doux2021,McCullough2024} both set fiducial scale cuts by requiring that the $\chi^2$ between a data vector predicted by their fiducial model and one with baryon contamination be less than 1. Ref.~\cite{Doux2021} also performs analyses with an alternative set of scale cuts intended to remove data that is sensitive to $k>k_{\rm max}$. These are very similar to the scale cuts employed in this analysis, and indeed their $k_{\rm max}=1\, \himpc$ scale cuts are nearly identical to our $\Lambda=1\, \himpc$ scale cuts when not using LCTs and so provide a useful point of comparison to our constraints. Ref.~\cite{Doux2021} does not report the best fit value for this analysis, hence the absence of a cross. We have not made an exhaustive attempt to reproduce the results in ref.~\cite{Doux2021}, although when using our version of the TATT model, and baryon counterterms without LCTs we find nearly identical mean $S_8$ values with $27\%$ larger uncertainties that are very likely attributable to the differences in our baryon and IA modeling choices, with potential differences also stemming from the updated redshift binning used in our measurements, correcting the binning used in the original DES-Y3 analyses. 

In order to investigate how our reported errors depend on these various modeling choices we perform a number of analysis variants. We display these variants in orange in Figure~\ref{fig:desy3_column}. While these are useful points of comparison we caution that these results come with additional assumptions beyond those made in our fiducial modeling choices. Analyses that fix the baryon counterterms to zero are labeled as ``no baryon model''. This analysis produces a mean and best fit that are nearly the same as our fiducial analysis, with $23\%$ smaller errors on $S_8$. We also conduct analyses using the Lagrangian perturbation theory version of the TATT and NLA models, which change the mean and best fit values only marginally while increasing the $S_8$ constraining power by $13\%$ and $30\%$ respectively. Finally, as an extreme case we also conduct an analysis that uses NLA while fixing our baryon counterterms to zero, which results in an increase in constraining power of $38\%$. 

\section{Conclusions}
\label{sec:conclusions}
Cosmic shear probes the cosmological clustering of matter and the growth of structure through the deflection of photons along the line-of-sight. Since galaxy shapes are distorted by the summed gravitational deflection from the source, the measured cosmic shear signal mixes contributions from a wide range of physical scales at a fixed angular scale, making the disentangling of large-scale structure and small-scale astrophysics difficult. This is a particular challenge because physical theories typically have well-defined scales of convergence beyond which their behavior quickly degrades, e.g. the nonlinear scale for perturbation theory or the baryonic scale for dark-matter only $N$-body simulations. 

In this paper, we propose a new treatment of the impact of small-scale (UV) matter clustering on cosmic shear observables through lensing counterterms (LCTs) inspired by effective field theory (EFT). Since weak lensing measured from different source samples probe the same underlying matter distribution, and the line-of-sight weighting is exactly known given the source redshift distribution, the UV dependence of cosmic shear 2-point functions can be summarized via a set of universal lensing counterterms encoding the scale and time dependence of the small-scale matter power spectrum. These LCTs are multiplied by the Taylor coefficients of the lensing kernel of each sample, which can be easily computed in terms of the mean inverse comoving distance of the sample. The total contribution to cosmic shear power spectra takes the form of a power series in the angular multipoles $\ell$. Testing this expansion in the case where the matter power spectrum and its time evolution are exactly given by their \texttt{halofit} predictions, we show that this power series rapidly converges over a wide range of $\ell$. In other words, the contributions of small-scale clustering to weak lensing observables on large angular scales is not arbitrary but rather described by smooth, polynomial shapes with coherent amplitudes across different source samples.

The LCT expansion allows for a consistent marginalization of the effects of unknown small-scale clustering on weak lensing. In the most straightforward case, one can marginalize over them allowing the small scale power spectrum, and therefore also the LCTs, to vary by e.g. $50\%$ or $100\%$ compared to some fiducial power spectrum like the dark-matter only power spectrum. When these priors are fixed independently of cosmology, marginalizing over these shapes is equivalent to adding a theoretical component to the covariance matrix. We test the LCT scheme on DES-Y3 and LSST-Y10-like mock data vectors and covariances, exploring its applicability for different values of the physical cutoff, order of the expansion and prior choices for the LCTs. For our fiducial choice of $\Lambda=1\himpc$, motivated by where our baryon counterterm model breaks down, we improve the $S_8$ precision over an analysis without lensing counterterms by $25\%$ for a DES-Y3-like analysis and $82\%$ for an LSST-Y10-like analysis. The large gain seen in the LSST-Y10-like analysis is somewhat driven by our requirement that contamination from $k>\Lambda$ be below the $1\%$ level in this case, compared to $5\%$ for the DES-Y3-like case. Figure~\ref{fig:kmax_sens} explores these trends in detail, including in the case when baryons are included. Figure~\ref{fig:lct_prior_sens} demonstrates that these improvements are largely not due to our choice of priors on the LCTs. Figure~\ref{fig:kmax_sens} then shows how these improvements depend on the assumed value for $\Lambda$ and assumptions about baryons, where the LSST-Y10 improvement decreases with increasing $\Lambda$ while the DES-Y3 improvement remains roughly constant.

The LCT scheme is similar in spirit to so-called nulling techniques in the literature which seek to reduce the UV dependence directly at the level of the observable \cite{Huterer05,Piccirilli25}, except that our method acts instead at the level of the model and likelihood. Nulling techniques have the advantage of being simple to implement and directly removing signal from UV contributions from the data, but are relatively blunt in their ability to remove UV information. In particular, because of the restriction to using linear combinations of the measured source bins, significant information from $k<\Lambda$ must be removed in order to ensure that the $k>\Lambda$ signal is nulled. Some $k<\Lambda$ information is also removed due to the relatively slow convergence of the LCT expansions, but this can be made significantly smaller if desired as outlined in Appendix~\ref{app:nonzero_pz}. When using $N=4$, as done throughout this work, we only lose $\sim 50\%$ of the $k<1\himpc$ constraining power on $S_8$ as shown in Figure~\ref{fig:kmax_sens}, compared to the over $200\%$ decrease in constraining power when removing the most contaminated mode when applying nulling with $k_{\rm max}=1\himpc$ reported in \cite{Piccirilli25}. We leave a more in depth investigation of the differences between our LCT methodology and nulling to future work.

Another promising aspect of the LCT scheme, compared to schemes such as nulling, is that the LCTs contain information about small-scale matter clustering. If their value is constrained to be different than the dark-matter only $\Lambda$CDM prediction, that is an indication that matter is more or less clustered at small scales compared to expectations, i.e. the small-scale matter clustering can be interpreted as an additional scale-dependent ``noise'' in cosmic shear observables beyond the large-scale modes modeled whose level we can measure. In order to test the feasibility of this kind of analysis we consider two separate cases where the small-scale power is (a) suppressed due to dark matter being an ultralight axion and (b) modulated by baryonic feedback as measured in the OWLS simulations. We find that the LCT expansion allows us to marginalize over both cases, removing large biases in some cases, and in addition detect the corresponding suppression of small-scale power while conventional analyses struggle to do either in these scenarios for a range of cutoff scales.

After validating our model on the Buzzard simulations, we analyze the public DES-Y3 cosmic shear data using the LCT scheme, using our own harmonic space measurements as a proof of principle. For our fiducial choice of $\Lambda=1\, \himpc$ we find $S_8 = 0.798\pm 0.026\, (0.826)$, improving over our constraint without LCTs of $S_8 = 0.780^{+0.035}_{-0.030}\, (0.797)$ by $19\%$, consistent with our findings on simulated data. Constraints on our baryon and intrinsic alignment parameters are only slightly improved. Our LCT methodology still allows for potential contributions to IA signals from $k>\Lambda$, and so some caution is required when using LCTs to improve IA constraints, a topic that we will return to in the near future. Our fiducial constraints prefer higher values of $S_8$ than previous analyses of the same shear catalogs in \cite{Amon22,Secco22,Doux2021,McCullough2024}, consistent with the $\Lambda$CDM results from the primary CMB \cite{Planck2018,Tristram2023}. Throughout this work, we make use of our \texttt{JAX}-based, machine-learning-accelerated analysis pipeline \texttt{gholax} that we make publicly available.

The LCT expansion described in this work can be applied to any lensing analysis that takes as input predictions of the nonlinear power spectrum, ranging from perturbative predictions from the EFT of large-scale structure to simulations-based predictions including baryonic effects. The former is an especially intriguing possibility, since it has a clearly predicted range of validity that evolves with redshift towards smaller scales---this type of cutoff can be easily incorporated into the LCT scheme. Beyond galaxy lensing, which we have focused on in this work, the LCTs apply equally to CMB lensing and joint galaxy-CMB lensing analyses. We intend to return to these topics in future work.

\section*{Acknowledgments}
SC acknowledges support
from the National Science Foundation at the IAS through
NSF/PHY 2207583. Support for this work was provided
by NASA through the NASA Hubble Fellowship grant
HST-HF2-51572.001 awarded by the Space Telescope
Science Institute, which is operated by the Association of
Universities for Research in Astronomy, Inc., for NASA,
under contract NAS5-26555. This work was performed in
part at the Aspen Center for Physics, which is supported
by National Science Foundation grant PHY-2210452.
SC thanks the Galileo Galilei Institute for Theoretical
Physics for the hospitality and the INFN for partial support during the completion of this work.

\appendix

\section{Time-Dependent Cutoff}
\label{app:time_dependent_kuv}
In this Appendix, we briefly consider the case of a time-dependent UV cutoff scale, i.e.
\begin{equation}
    P^{\rm UV}_{mm}(k,\chi) = P_{\rm mm}(k,\chi)\ \Theta(k - \Lambda(\chi)).
\end{equation}
This example is relevant for applying perturbation theory to cosmic shear, since $k_{\rm NL}$ moves quickly to smaller scales at higher redshift. In this case the derivatives of the power spectrum now involve delta functions, e.g.
\begin{align}
    \frac{\partial P^{\rm UV}_{mm}}{\partial \chi} = &\frac{\partial P_{\rm mm}}{\partial \chi} \Theta(k - \Lambda(\chi)) \nonumber \\
    &- P_{\rm mm}(k,\chi) \left( \frac{d \Lambda}{d \chi} \right) \delta_D(k -\Lambda(\chi)).
\end{align}
These delta functions lead to boundary terms in $\sigma_{N,o}^{\rm UV}$. The first one appears at $N=3$ with
\begin{align}
    C^{ab}_{\ell, \rm UV} \supset - &\left(\frac32 \Omega_m H_0^2 \right)^2 \left( \ell + \frac12 \right)^2  \nonumber \\
    &\quad \left( \Lambda^{-3} \frac{d \Lambda}{d \chi} \right) P_{\rm mm}^{\rm UV}(\Lambda_{0},z=0)  .
\end{align}
The negative sign of this contribution is expected---accounting for better ranges of fit at higher, more nonlinear scales reduces the contributions of small-scale clustering since more scales are now model-able. Importantly, changing the functional form of the UV contribution in this way does not change the functional form in Equation~\ref{eqn:lensing_ct_expansion} but only the interpretation of its coefficients.

\section{LCT Expansion without Low-Redshift Source Suppression}
\label{app:nonzero_pz}

\begin{figure}[t!]
    \includegraphics[width=0.5\textwidth]{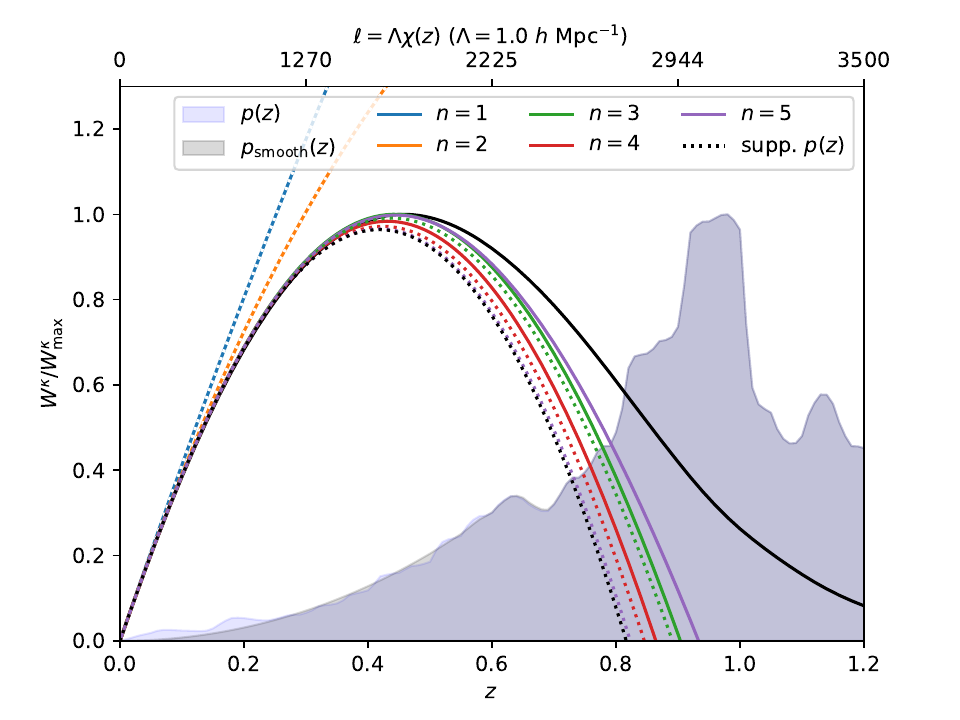}
    \caption{Convergence of the series for $W^{\kappa^i}$ at order $\chi^n$ when the low-$z$ behavior of the source distribution is taken into account (solid) and assumed to be extremely suppressed (dotted). The two series are equivalent at linear and quadratic order. The black dashed line shows the $n=\infty$ prediction in the latter case, given by Equation~\ref{eqn:Wchi_smallp}. The (smoothed) redshift distribution (used to estimate its derivatives) is shown in shaded blue and gray. An upper panel shows the corresponding angular multipole at a given redshift subtended by the cutoff scale $\Lambda^{-1}$.}
    \label{fig:wmu_convergence}
\end{figure}

Following Equation~\ref{eqn:lensing_kernel_theta}, the source distribution dependent part of the lensing kernel $f^a$, where we have defined $W^{\kappa^i} = A (1 + z)\chi f^i$, is
\begin{align}
    f^i(\chi) &= \int_0^\infty d\chi'\ p^i_\chi(\chi') \left( \frac{\chi' - \chi}{\chi'} \right) \Theta(\chi' - \chi).
\end{align}
We are interested in the derivatives of this function as $\chi \rightarrow 0^+$. The first derivative is given by
\begin{align}
    &\frac{df^i}{d \chi} = \int_0^\infty d\chi'\ p^i_\chi(\chi') \Big( -\frac{\Theta(\chi'-\chi)}{\chi'} + \nonumber \\
    &\Big( \frac{\chi' - \chi}{\chi'} \Big) \delta_D(\chi'-\chi) \Big) = - \left \langle \frac{\Theta(\chi'-\chi)}{\chi'}  \right \rangle \xrightarrow{\chi \rightarrow 0^+} \langle \chi^{-1} \rangle_i.
\end{align}
The limit is well-defined if $p_\chi(\chi) \propto \chi^n$ for $n > 0$ at small $\chi$ or, when $p_\chi$ can be Taylor expanded, $p_\chi(0) = 0$, i.e. the distribution is at most linear in $\chi$.
A similar calculation gives
\begin{equation}
    \frac{d^2 f^i}{d\chi^2} = \frac{p^i_\chi(\chi)}{\chi} \xrightarrow{\chi \rightarrow 0^+} p_\chi^{i (1)}(0).
\end{equation}
Indeed, for analytic $p^i_\chi$ we have that generically $d^{2+n} f/d\chi^{2+n} = p^{(n+1)}(0)/(n+1)$ as long as $p(0) = 0$, i.e. the lensing kernel is smooth as long as the distribution is at most linear at low $\chi$.

Figure~\ref{fig:wmu_convergence} shows the convergence of the Taylor series of $W^{\kappa^i}$ with and without our fiducial assumption that $p^{i}(z)$ is extremely suppressed near the origin.  In the former case, the series converges to Equation~\ref{eqn:Wchi_smallp}, which differs from the true lensing kernel at high $z$, though both give good approximations to the lensing kernel at low $z$ at angular scales close to our chosen cutoff $\Lambda$. Note that the two expansions coincide at the linear and quadratic orders since $f^i(0) = 1$ and $f'^{i}(0) = \langle \chi^{-1} \rangle_i $ are independent of the derivatives of the redshift distribution at the origin. In order to compute the full expansion we have approximated the low-$z$ behavior of the source distribution as a smooth function, as shown in the figure.

\section{Alternative Parametrization}
\label{app:alt_expansion}
It is also possible to rewrite the expansion in terms of
\begin{equation}
    \tilde{W}^{\kappa^i}(\chi) = \frac{W^{\kappa^i}(\chi)}{1 + z(\chi)}, \, \tilde{P}_{mm}(k,z) = (1+z)^2 P_{\rm mm}(k,z).
\end{equation}
In this case the modified coefficients are
\begin{equation}
    \tilde{w}_1^i = 1, \, \tilde{w}_2^i = - \langle \chi^{-1} \rangle_i, \, \tilde{w}_{n>2}^i = 0.
\end{equation}
In an Einstein-de Sitter universe, the linear growth factor is $D(z) = (1+z)^{-1}$, such that the re-scaled $\tilde{P}_{mm}$ has a lower dynamical range, leading to a potentially more well-converged expansion about $\chi = 0$. However, in $\Lambda$CDM universes close to our own, the influence of dark energy near $z = 0$ makes this rescaling somewhat sub-optimal.

\section{Emulator Construction}
\label{app:emu_construction}
Throughout this work we make use of neural network emulators to accelerate the calculation of our intrinsic alignment and matter power spectrum models. We largely use the same methodology as described in \cite{Chen24b}, which built on the work in \cite{DeRose2021}, with a few minor changes. Most importantly, we predict spectra that are normalized by $D(z)^2$ to remove the dominant redshift evolution of the spectra. We then train an emulator to predict $D(z)^2$ alone, and rescale the emulator predictions by this at prediction time. This significantly increases the accuracy of our emulators at a fixed number of training points. 

For our intrinsic alignment model, we train an emulator to predict basis spectra entering into $P^{(m)}_{22}(k,z)$ and $P^{(0)}_{02}(k,z)$ using one neural network each. This has a minor computational speed advantage, speeding up the evaluation time by about a factor of 13 and 21 i.e. the number of basis spectra that enter into $P^{(m)}_{22}(k,z)$ and $P^{(0)}_{02}(k,z)$ respectively. The Limber integration step of our likelihood is the bottleneck of our calculations, so these speedups are not particularly important. There is also a commensurate decrease in the memory footprint of these emulators, which is more significant as the size of these emulators is the dominant contribution to the size of our likelihood code in RAM.

\bibliography{main}

\end{document}